\documentclass[acmtog]{acmart} 
\acmSubmissionID{0527}

\usepackage{booktabs} 
\usepackage{amsmath}

\usepackage{amsthm}
\usepackage{lipsum}
\usepackage{mathtools}
\usepackage{siunitx}
\usepackage{hyperref}
\usepackage{subcaption}
\usepackage{afterpage}
\usepackage{multirow}
\usepackage{overpic}
\usepackage{wrapfig}
\usepackage{multirow}
\usepackage{graphicx}
\usepackage{soul}
\usepackage{caption}

\usepackage{wrapfig}

\newcommand{\R}{\mathbb{R}}
\let \bs=\mathbf
\let \set=\mathcal

\def \E {\mathbb{E}}

\def \diag {\mathrm{diag}}

\def \KL {\textup{KL}}

\def \path {\mathit{path}}

\let \set = \mathcal
\let \bs = \boldsymbol

\usepackage[ruled]{algorithm2e} 
\usepackage{algorithmic}
\usepackage[export]{adjustbox}

\SetAlFnt{\small}
\SetAlCapFnt{\small}
\SetAlCapNameFnt{\small}
\SetAlCapHSkip{0pt}

\acmJournal{TOG}

\definecolor{georgecolor}{RGB}{255, 87, 51}

\hypersetup{final}

\begin{document}

\title{GenAnalysis: Joint Shape Analysis by Learning Man-Made Shape Generators with Deformation Regularizations}

\author{Yuezhi Yang}
\affiliation{
\institution{The University of Texas at Austin}
\streetaddress{2317 Speedway}
\city{Austin}
\state{Texas}
\postcode{78712}
\country{USA}
}
\email{yyuezhi123@gmail.com}

\author{Haitao Yang}
\affiliation{
\institution{The University of Texas at Austin}
\streetaddress{2317 Speedway}
\city{Austin}
\state{Texas}
\postcode{78712}
\country{USA}
}
\email{yanghtr@outlook.com}

\author{Kiyohiroe Nakayama}
\affiliation{
\institution{Stanford University}
\city{Palo Alto}
\state{California}
\country{USA}
}
\email{nakayama4756677@gmail.com}

\author{Xiangru Huang}
\affiliation{
\institution{Westlake University}
\city{Palo Alto}
\state{California}
\country{China}
}
\email{xiangruhuang816@gmail.com}

\author{Leonidas Guibas}
\affiliation{
\institution{Stanford University}
\city{Palo Alto}
\state{California}
\country{USA}
}
\email{guibas@cs.stanford.edu}

\author{Qixing Huang}
\affiliation{
\institution{The University of Texas at Austin}
\streetaddress{2317 Speedway}
\city{Austin}
\state{Texas}
\postcode{78712}
\country{USA}
}
\email{huangqx@cs.utexas.edu}

\begin{abstract}

We present GenAnalysis, an implicit shape generation framework that allows joint analysis of man-made shapes, including shape matching and joint shape segmentation. The key idea is to enforce an as-affine-as-possible (AAAP) deformation between synthetic shapes of the implicit generator that are close to each other in the latent space, which we achieve by designing a regularization loss. It allows us to understand the shape variation of each shape in the context of neighboring shapes and also offers structure-preserving interpolations between the input shapes. We show how to extract these shape variations by recovering piecewise affine vector fields in the tangent space of each shape. These vector fields provide single-shape segmentation cues. We then derive shape correspondences by iteratively propagating AAAP deformations across a sequence of intermediate shapes. These correspondences are then used to aggregate single-shape segmentation cues into consistent segmentations. We conduct experiments on the ShapeNet dataset to show superior performance in shape matching and joint shape segmentation over previous methods.

\end{abstract}

\begin{CCSXML}
<ccs2012>
<concept>
<concept_id>10010147.10010371.10010396.10010402</concept_id>
<concept_desc>Computing methodologies~Shape analysis</concept_desc>
<concept_significance>500</concept_significance>
</concept>
</ccs2012>
\end{CCSXML}

\ccsdesc[500]{Computing methodologies~Shape analysis}
\ccsdesc[500]{Neural networks}

\keywords{shape correspondence, joint shape segmentation, neural network regularization}

\maketitle

\section{Introduction}

Shape analysis is a fundamental research area in geometry processing that enjoys many applications~\cite{Mitra:2014:SIG,XuKHK17}. Existing shape analysis approaches fall into the categories of supervised and unsupervised methods. Although supervised methods achieved state-of-the-art results, they require user labels, which are costly to obtain and do not scale up. In contrast, unsupervised shape analysis methods~\cite{Deng2021DIF-Net,Shuai2023DPF-Net,Chen2024DAE-Net} benefit from learning from large-scale unlabeled shapes and have recently shown promising results in the large-scale ShapeNetPart~\cite{10.1145/2980179.2980238} dataset. 

\begin{figure}[t]
\centering
\begin{overpic}[width=1.0\columnwidth]{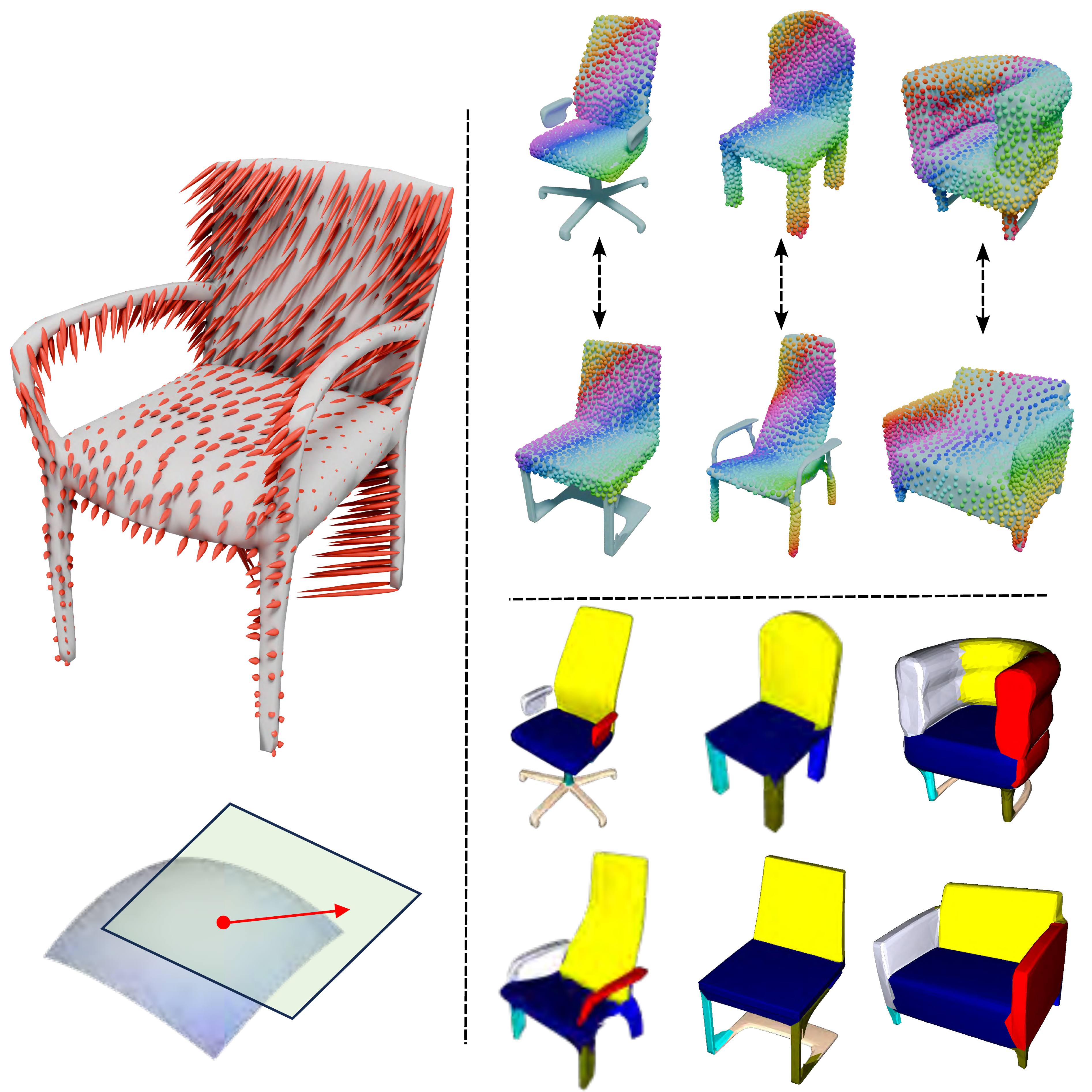}
\put(56, -1){consistent segmentation}
\put(60, 46){shape matching}
\put(-3, -1){as-affine-as-possible deformation}
\end{overpic}
\caption{GenAnalysis learns a shape manifold from a collection of man-made shapes with a novel as-affine-as-possible deformation regularization loss. The learned shape manifold supports shape analysis by analyzing the tangent space of each shape, shape matching through intermediate shapes on this manifold, and consistent segmentation by aggregating single-shape analysis results using inter-shape correspondences. }
\label{Figure:Teaser}
\end{figure}

The fundamental challenge in developing unsupervised techniques is to accommodate significant structural and geometric variations across the input shapes. A popular paradigm is learning template-based models in which each template consists of a bag of (non-rigidly) deforming parts. The learning phase learns parameters of the templates to fit the training data. These include early works of~\cite{DBLP:journals/tog/OvsjanikovLGM11,DBLP:journals/tog/KimLMCDF13,akzm_shapeSynth_eg14,DBLP:journals/cgf/HuangKM15,Huang:2015:ASS} and recent deep learning methods~\cite{DBLP:conf/3dim/Ganapathi-Subramanian18,DBLP:conf/iccv/GenovaCVSFF19,deprelle2019atlasV2,Shuai2023DPF-Net,Chen2024DAE-Net,Zheng2021DIT-Net,Deng2021DIF-Net}. The key advantage of template-based models is that during the testing phase it is only required to fit template to each shape, which is efficient and simple. However, template-based models exhibit two major limitations. First, template-based models rely on strong prior knowledge of shape parts (e.g., the number of parts and their spatial locations) either explicitly or implicitly. Second, the learning and inference phases can easily get stuck in local minimums.

In this paper, we introduce a novel shape analysis framework, named GenAnalysis, which learns an implicit shape generative model to fit the training shapes. We will show how to use this template-free shape generative model to perform shape analysis, hence circumventing the limitation of template-based models. Our approach is inspired by the success of studying organic shape collections using the lens of shape spaces as differential manifolds~\cite{DBLP:journals/pami/KlassenSMJ03,Kilian:2007:MSS,DBLP:journals/pami/SrivastavaKJJ11,DBLP:journals/ijcv/HartmanSKCB23,10.1145/3618371}. Moreover, recent advances in shape generative models~\cite{ParkFSNL19,DBLP:conf/cvpr/ChenZ19,DBLP:conf/cvpr/MeschederONNG19,DBLP:conf/iclr/AtzmonL21,Yang:2023:ICCV} can accurately fit man-made training shapes that exhibit much larger geometric variations than organic shapes. These learned continuous man-made shape spaces offer novel means of shape analysis. Specifically, they provide smooth interpolations between two shapes that can be used to compute shape correspondences. 
Moreover, when trained properly,these generative models enable us to understand the shape variations among the shape collection for shape segmentation by analyzing shape variations in the tangent space of each shape when viewing the learned generative models as shape manifolds.

However, learning a shape generative model is under-constrained, as there are many possible network parameters and latent codes of training shapes that fit the training data. Learning shape generative models under generic distribution alignment paradigms~\cite{10.1145/3422622,DBLP:journals/corr/KingmaW13,10.1145/3592442} do not result in shape generators that offer meaningful intermediate shapes and tangent spaces for shape matching and shape segmentation. Unlike deformable objects in which isometric deformations offer faithful deformation priors, the structural shape variations in man-made shapes are too complex to be summarized into a concise universal principle. The suitable regularization losses are heavily dependent on the application scenarios.

GenAnalysis introduces a novel regularization loss for learning the shape generator. This loss is tailored for shape analysis.

\begin{figure}[t]
  \centering
  
  \begin{overpic}[
      width=\columnwidth,
      trim=0pt 12pt 0pt 18pt,  
      clip                     
    ]{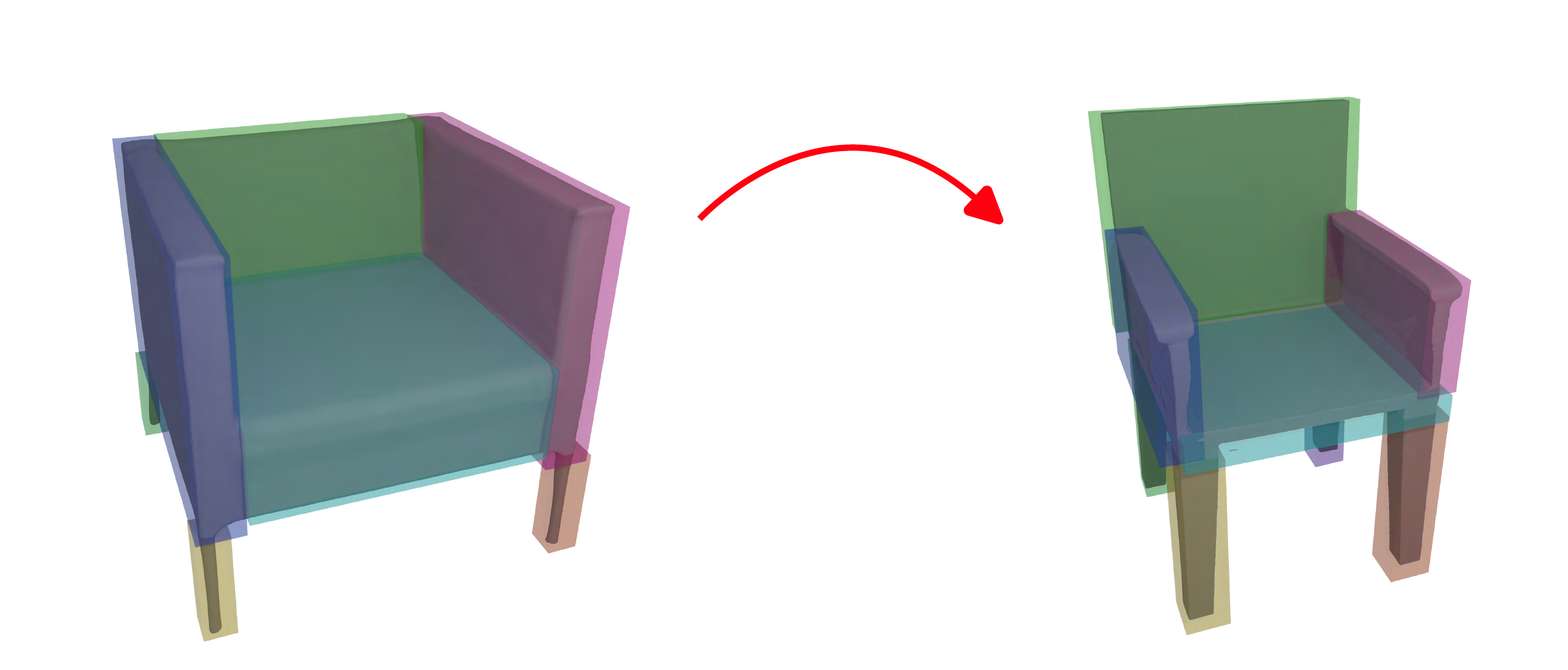}
    \put(51,28){\textcolor[rgb]{1,0,0}{$A_i$}}
  \end{overpic}

  \captionsetup{aboveskip=2pt, belowskip=-4pt}

  \caption{\textbf{Piece-wise affine assumption.}
    Each shape part from the source shape, approximated by its bounding
    box, undergoes an affine transformation $A_i$ to the corresponding
    part in the target shape.}
  \Description{Two chair silhouettes divided into parts; arrows show
    each part mapped by its own affine transform $A_i$.}
  \label{Figure:piece-wise-affine}
\end{figure}

The proposed loss builds on the popular piece-wise affine assumption~\cite{10.1145/1866158.1866206,DBLP:journals/tog/OvsjanikovLGM11,DBLP:journals/tog/KimLMCDF13,DBLP:journals/cgf/HuangKM15,Chen2024DAE-Net}, in which \textbf{if we abstract each shape part using a bounding box, then the deformations at the part level are affine}  (see Figure~\ref{Figure:piece-wise-affine}). It also enforces local piece-wise affine assumption in a more flexible and faithful way compared to template based approaches hence alleviates the local minimum issue by achieving better shape quality.
However, the challenge is how to model the significant shape variations within each part. GenAnalysis presents two strategies to address this challenge. First, we minimize the piece-wise affine deformation between adjacent synthesized shapes defined by the generator (i.e., they are close in the latent space). This approach is analogous to the smoothness loss in curve fitting, but we model strong prior knowledge about inter-shape deformations. In particular, we show how to employ an as-affine-as-possible deformation loss under a robust norm to fulfill the piece-wise affine assumption. 
Second, we describe a weighting scheme for test-time optimization~\cite{
DBLP:conf/iclr/WangSLOD21,DBLP:conf/cvpr/0001WDE22,DBLP:conf/icml/NiuW0CZZT22,DBLP:journals/corr/abs-2307-05014}. It prioritizes that the regularization loss distributes deviations of piece-wise affine deformations away from shapes of interest. This allows us to extract segmentation cues by analyzing the tangent space of each shape (i.e., vector fields that characterize differences between each shape and its adjacent shapes in the latent space).

We have evaluated the performance of GenAnalysis on ShapeNet for the task of shape matching and consistent shape segmentation. In both tasks, GenAnalysis outperforms state-of-the-art techniques by salient margins, justifying the power of performing shape analysis by learning shape generators. For example, GenAnalysis outperforms DAE-Net~\cite{Chen2024DAE-Net} by 3.2\% in mean IOU score. An ablation study justifies the importance of each component of GenAnalysis.

\section{Related Work}
\label{Section:Related:Works}

\subsection{Neural Generative Models for Man-Made Shapes}
\label{Subsec:Neural:Generative:Models}

Recent advances in learning generative shape models have shown impressive results in shape synthesis. These approaches have focused on developing network architectures under different 3D representations~\cite{AchlioptasDMG18,NIPS2016_44f683a8,ParkFSNL19,Li2017GRASS,2019structurenet, Siddiqui2023MeshGPTGT, zhang2024clay}. A common scheme is to align the distribution of the training shapes with that of the synthetic shapes~\cite{ArjovskyCB17,10.1145/3422622,LitanyBBM18,ranjan2018generating,bouritsas2019neural,YangLH18,ParkFSNL19,zadeh2019variational,DBLP:conf/nips/SongE19,DBLP:conf/iclr/0011SKKEP21}. However, these approaches are limited when the training shapes are sparse. An approach to address this data issue is to develop regularization losses that enforce prior knowledge about the underlying shape space~\cite{DBLP:conf/iccv/HuangHSZJB21,DBLP:conf/icml/GroppYHAL20,DBLP:journals/corr/abs-2108-08931,10.1145/3528233.3530713,DBLP:conf/cvpr/MuralikrishnanC22,Yang:2023:ICCV}. The novelties of GenAnalysis are addressing man-made shapes that have large shape variations and learning the shape generator to promote simple extractions of shape segmentation cues in the tangent space of each shape. 

\subsection{Data-Driven Shape Segmentation}
\label{Subsec:Joint:Shape:Seg}

A common paradigm among non-deep learning-based joint shape segmentation approaches~\cite{Golovinskiy:2009:CSO,10.1145/2070781.2024159,DBLP:journals/tog/SidiKKZC11,10.1111/j.1467-8659.2012.03175.x,DBLP:conf/iccv/WangHG13,DBLP:conf/cvpr/WangHOG14,Huang:2014:FMN,Huang:2019:TM} is to promote the segmentation of each shape to be consistent in a shape collection, using point-wise correspondences~\cite{10.1145/2070781.2024159}, primitive-level correspondences~\cite{10.1145/2070781.2024159,DBLP:journals/tog/SidiKKZC11,10.1111/j.1467-8659.2012.03175.x}, or functional maps~\cite{Ovsjanikov:2012:FMF,DBLP:conf/iccv/WangHG13,DBLP:conf/cvpr/WangHOG14,Huang:2014:FMN,Huang:2019:TM}. However, these approaches only exhibit limited performance in heterogeneous shape collections, where computing consistent correspondences between structurally dissimilar shapes is challenging.

Deep learning based methods are introduced to handle joint shape segmentation for its improved generalization ability on heterogeneous shape collection. Most of the methods for learning consistent segmentation are supervised learning~\cite{Charles2017pointNet,Charles2017pointNet++,zhao2019cpsule,evangelos2017projCNN}. To alleviate the need for 3D annotation, unsupervised methods learn deformable part templates to reconstruct input shapes for shape segmentation. These approaches use a variety of part abstractions, including cuboids~\cite{Tulsiani2017VP,Sun2019HA,Yang2021CA}, superquadratics~\cite{Paschalidou2019SQ}, part point clouds~\cite{DBLP:journals/cgf/HuangKM15}, convex polygons~\cite{Deng2020Cvx-Net,Chen2020BSP-Net}, sphere templates~\cite{Paschalidou2021NeuralParts} and implicit templates ~\cite{Chen2019BAE-Net,Niu2022RIM-Net,Chen2024DAE-Net}. Part deformations can also be encoded explicitly~\cite{DBLP:journals/cgf/HuangKM15} or implicitly using neural network branches~\cite{Chen2019BAE-Net,Shuai2023DPF-Net,Paschalidou2021NeuralParts,Tertikas2023PartMerf,Chen2024DAE-Net}. However, they still require some prior knowledge of the underlying part structure, e.g., the number of parts. The designed template often exhibits limited performance when expressing complex variations in each shape part. GenAnalysis circumvents the limitations of template-based approaches in two ways. First, as we will discuss next, GenAnalysis presents dense shape correspondences between shapes with large structural variations to aggregate shape segmentation cues. Second, GenAnalysis enforces the prior knowledge about shape parts using a regularization loss. The regularization loss is enforced on adjacent synthetic shapes and tolerates large inter-shape variations.

Several approaches, including STAR~\cite{abdelreheem2023satr}, 3D Highlighter~\cite{decatur20233dhighlighter}, and follow-up methods~\cite{lang2024iseg,yang2023sam3d,cen2023segmentNerf,kontogianni2023interactiveseg},  have explored 3D segmentation using vision-language  models (VLM)~\cite{alayrac2022flamingo,zhang2022glipv2, ravi2024sam2}. The key idea is to lift 2D features from off-the-shelf CLIP \cite{radford2021clip}, GLIP \cite{zhang2022glipv2,li2022glip}, and SAM \cite{kirillov2023sam} models. One challenge of these approaches is to enforce multi-view consistency and address occlusions. Complementary to lifting 2D features, GenAnalysis is a 3D approach that focuses on consistent segmentation using a shape generative model.

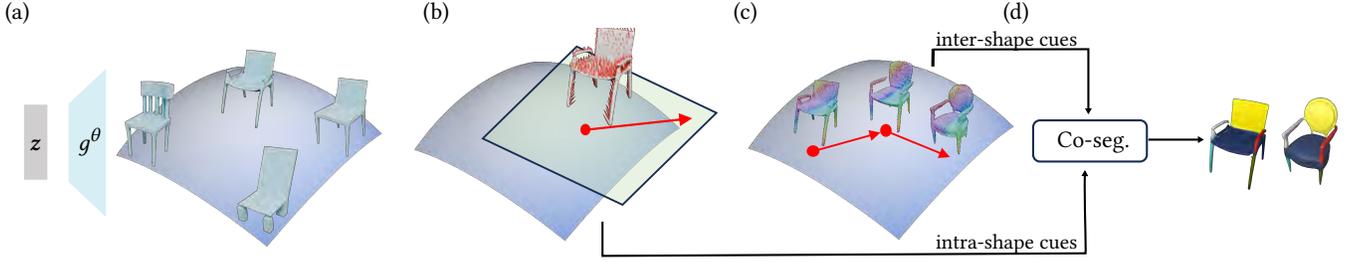
\begin{figure*}
\begin{overpic}[width=1.0\linewidth]{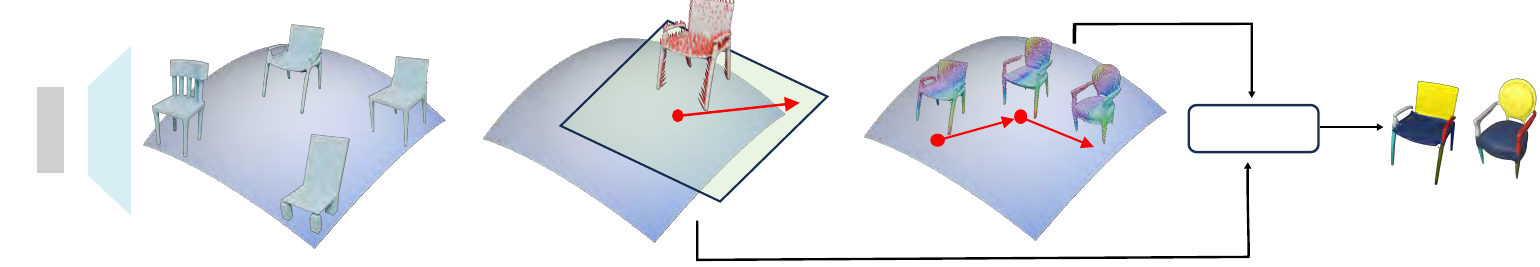}
\put(2.85, 8.3){$\bs{z}$}
\put(6.5, 8.3){$g^{\theta}$}
\put(1, 18){(a)}
\put(32, 18){(b)}
\put(55, 18){(c)}
\put(75, 18){(d)}
\put(70, 1){\small intra-shape cues}
\put(70, 16){\small inter-shape cues}
\put(79, 8.5){Co-seg.}
\end{overpic}
\caption{\textbf{The pipeline of GenAnalysis pipeline}, which consists of four stages. (a) The first stage learns an implicit shape generator to fit the input shapes by combing a data loss and an as-affine-as possible deformation loss.  (b) The second stage extracts piece-wise affine structures in vector fields of each shape derived from the tangent space of each shape. (c) The third stage computes pairwise shape correspondences obtained by composing correspondences along intermediate shapes defined by the generator. Points with similar color are in correspondence. (d) The last stage performs consistent segmentation using the correspondences obtained in stage three to integrate single-shape segmentation cues derived from stage two. }
\label{Figure:Overview2}
\end{figure*}

\subsection{Shape Matching and Deformation Models}
\label{Subsec:Inter:Shape:Corres}

Computing correspondences between geometric shapes is a long-standing problem in computer graphics. It is beyond the scope of this paper to provide a comprehensive review. We refer to~\cite{DBLP:conf/eurographics/Kaick0HC10,DBLP:journals/tvcg/TamCLLLMMSR13,DBLP:journals/vc/Sahillioglu20} for surveys on this topic. Most approaches~\cite{10.1145/2010324.1964974,10.5555/3157382.3157455,10.1145/3355089.3356524} deal with deformable shapes assume that they have consistent topologies and full dense correspondences are well-defined. Such approaches do not apply to man-made shapes with complex structure variations where dense correspondences are partially defined.

A closely related problem to correspondence computation is non-rigid shape registration~\cite{Tam:2013:RPC}, which deforms a source shape under some deformation model to match a target shape. In this context, the deformation model plays a key role. The most widely used deformation models are as-rigid-as-possible (ARAP)~\cite{Alexa:2000:ARAP,Sorkine:2007:ARM,DBLP:conf/iccv/HuangHSZJB21} and as-conformal-as-possible (ACAP)~\cite{DBLP:journals/cgf/YoshiyasuMYK14,10.1145/3618371} models. These models are mostly used for organic shapes but not for man-made shapes that exhibit sophisticated part deformations. For man-made shapes, a widely used assumption is piece-wise affine, which has been used in template-based models, that consists of rectangular bounding boxes. GenAnalysis shows how to apply the idea of non-rigid shape registration to find correspondences across man-made shapes. This is achieved by using a robust as-affine-as-possible deformation model to enforce the piece-wise affine deformation and performing non-rigid registrations between adjacent synthetic shapes of a man-made shape generator. The idea of optimizing an interpolation between two shapes for shape matching has also been explored in NeuroMorph \cite{eisenberger2021neuromorph} and GenCorres \cite{Yang:2023:ICCV}, both of which study organic shapes. In contrast, GenAnalysis focuses on man-made shapes which have much more complex geometric variations, where the deformation models of NeuroMorph and GenCorres do not apply.

In contrast to template-based approaches~\cite{Deng2021DIF-Net,Zheng2021DIT-Net,Kim2021SemanticDIF-Net,Kyle2019SIF} for made-made shapes, GenAnalysis exhibits improved performance on heterogeneous shape collections. This is because template-based models may not be expressive enough and/or the learning procedure can get trumped into local minimums.

\section{Problem Statement and Approach Overview}
\label{Section:Overview}

This section presents the problem statement (Section~\ref{Subsec:Problem:Statement}) and an overview of GenAnalysis (Section~\ref{Subsec:Approach:Overview}).

\subsection{Problem Statement}
\label{Subsec:Problem:Statement}

The input to GenAnalysis is a collection of training man-made shapes $\set{S} = \{S_i\} \subset \overline{\set{S}}$ where $\overline{\set{S}}$ denotes the ambient shape space. GenAnalysis aims to learn a generative model that enables two fundamental shape analysis tasks on a collection of test shapes $\set{S}_{\textup{test}}$: 
\begin{itemize}
\item \textsl{Joint Shape Segmentation.} For each test shape $S\in \set{S}_{\textup{test}}$, we want to decompose it into parts based on the underlying shape variations provided by $\overline{\set{S}}$. 
\item \textsl{Shape Correspondence.} Given a test source shape $S \in \set{S}_{\textup{test}}$, a test target shape $S' \in \set{S}_{\textup{test}}$, and a point $\bs{p}$ in $S$, we want to calculate its correspondence in $S'$. The output indicates whether this correspondence is well defined (due to partial similarities) and, if so, the corresponding point. 
\end{itemize}

\subsection{Approach Overview}
\label{Subsec:Approach:Overview}

GenAnalysis proceeds in four stages (see Figure~\ref{Figure:Overview2} for an illustration).

\subsubsection{Shape Generator Learning}
The first stage learns an implicit shape generator from $\set{S}$. We introduce a novel regularization loss to enforce piece-wise affine deformations between synthetic shapes that are close to each other in the latent space. The loss employs an as-affine-as-possible (AAAP) deformation model under the $L^2$ norm for correspondence computation and the same AAAP model under a robust norm to enforce piece-wise affine deformations. Network training integrates this regularization loss into a standard auto-decoder paradigm. In addition, we introduce a light-weight test-time optimization step to improve the alignment between the generative model and test shapes and distribute distortions of the piece-wise affine assumption away from the tangent spaces at test shapes. This weighting scheme facilitates the analysis of the tangent space at each test shape to extract shape segmentation cues.

\subsubsection{Shape Variation Analysis}
The second stage analyzes the tangent spaces of the learned shape manifold. The tangent space of each shape encodes the shape variations in its infinitesimal neighboring shape space. The regularization loss employed in stage one prioritizes that this tangent space contains variation vector fields that exhibit piece-wise affine structures. We present a spectral approach to extract such vector fields and show how to analyze their underlying piece-wise affine structures to derive a distance matrix, in which the distance between two points is small if they belong to the same part and vice versa. Figure 8 (Left) shows some examples of the extracted vector fields.

\subsubsection{Shape Matching} 
The third stage of GenAnalysis computes shape correspondences between pairs of test shapes. We show that unlike matching pairs of shapes directly, which leads to poor results under large shape variations, GenAnalysis propagates correspondences derived from AAAP deformation between intermediate shapes interpolated from the learned deformation model. The propagation procedure combines displacement and projection operations to ensure that the propagated correspondences lie on intermediate shapes. The resulting correspondences will also be used to aggregate single-shape segmentation cues from the second stage.

\subsubsection{Consistent Segmentation}
The last stage of GenAnalysis performs a consistent segmentation. We formulate a spectral-based consistent segmentation approach that integrates the single-shape segmentations derived from shape variation analysis using inter-shape correspondences.

\section{Approach}

This procedure describes four stages of GenAnalysis in detail (from Section~\ref{Subsec:Shape:Generator} to Section~\ref{Subsec:Cons:Seg}).

\subsection{Shape Generator Learning}
\label{Subsec:Shape:Generator}

\begin{figure*}
\begin{overpic}[width=1.0\textwidth]{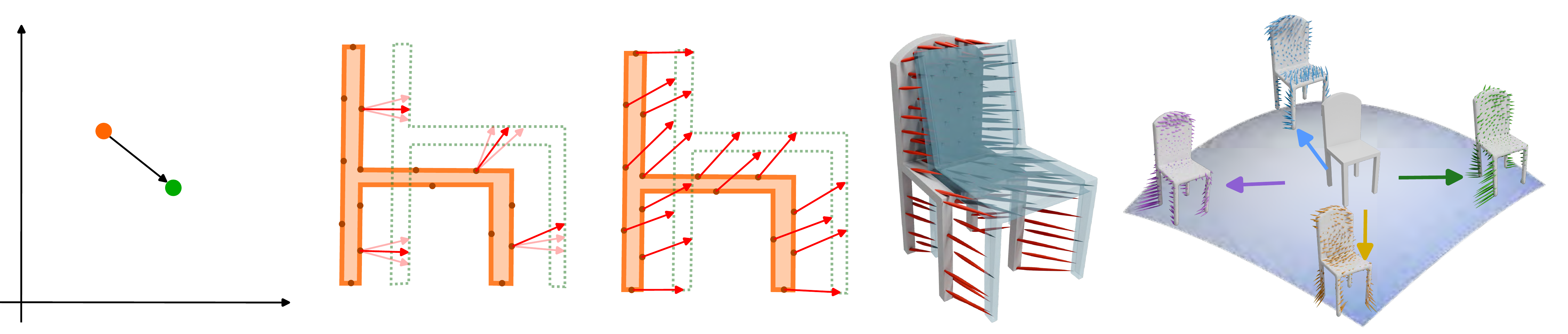}
\put(9, -2){(a)}
\put(27, -2){(b)}
\put(45, -2){(c)}
\put(63, -2){(d)}
\put(86, -2){(e)}
\put(0.5, 20){$\set{Z}$}
\put(6, 11){\textcolor[rgb]{1.0,0.4,0}{$\bs{z}$}}
\put(7.5, 9.5){$\bs{v}$}
\put(10, 7.5){\textcolor[rgb]{0,0.666,0} {$\bs{z}+\epsilon\bs{v}$}}
\put(26.5, 14.5){\textcolor[rgb]{0,0.666,0} {\scriptsize $g^{\theta} (\bs{x},\bs{z}+\epsilon\bs{v}) = 0$}}
\put(23,1){\scriptsize $\textcolor[rgb]{0.666,0.2666,0}{\bs{p}_i^{\theta}(\bs{z})} + \epsilon\textcolor[rgb]{1.0,0,0}{\bs{d}_i^{\bs{v}}(\bs{z})}$}
\put(80.5, 8){\textcolor[rgb]{0.553,0.373,0.827}{$v_1$}}
\put(82, 11){\textcolor[rgb]{0.333,0.6,1}{$v_2$}}
\put(90, 8.5){\textcolor[rgb]{0,0.666,0}{$v_3$}}
\put(88, 6){\textcolor[rgb]{0.831,0.666,0}{$v_4$}}
\put(72, 5.5){$\textcolor[rgb]{0.553,0.373,0.827}{\bs{d}^{\bs{v_1}}(\bs{z})}$}
\put(84, 18){\textcolor[rgb]{0.333,0.6,1}{$\bs{d}^{\bs{v_2}}(\bs{z})$}}
\put(95, 6){\textcolor[rgb]{0,0.666,0}{$\bs{d}^{\bs{v_3}}(\bs{z})$}}
\put(82, 0){\textcolor[rgb]{0.831,0.666,0}{$\bs{d}^{\bs{v_4}}(\bs{z})$}}
\put(87.5, 8){$\bs{z}$}
\put(39, 1){\scriptsize $\textcolor[rgb]{0,0.666,0}{\bs{p}_i^{\theta}(\bs{z}+\epsilon\bs{v})}\approx \textcolor[rgb]{0.666,0.2666,0}{\bs{p}_i^{\theta}(\bs{z})} + \epsilon\textcolor[rgb]{1,0,0}{\bs{d}_i^{\bs{v}}(\bs{z})}$}
\end{overpic}
\caption{\textbf{As-affine-as-possible (AAAP) regularization.} (a) We study infinitesimal perturbation $\bs{v}$ in the tangent space at each shape with latent code $\bs{z}$. (b) Due to constraint shown in Eq.~(\ref{Eq:Implicit:Cons}), we can not determine the correspondence $\bs{d}_i^{\bs{v}}(\bs{z})$ that lies on $g^{\theta} (\bs{x},\bs{z}+\epsilon\bs{v}) = 0$ directly. (c) We instead jointly compute all $\bs{d}_i^{\bs{v}}(\bs{z})$ by solving an constrained optimization problem using the objective function in Eq.~(\ref{Eq:E:Obj}). (d) We show resulting 3D correspondences between source shape colored in white, and a neighboring perturbed shape colored in transparent blue. (e) After derivation in section~\ref{Subsubsec:AAAP:Deformation}, we arrived at closed form solution shown in Eq.~(\ref{Eq:d:Explicit:Expression}) where each perturbation $v_i$ corresponds to variation at $\bs{d}^{\bs{v_i}}(\bs{z})$. We integrate over all directions to obtain the regularization term in Eq.~(\ref{Eq:Geo:Regu}).}
\label{Figure:AAAP:Detail}
\end{figure*}
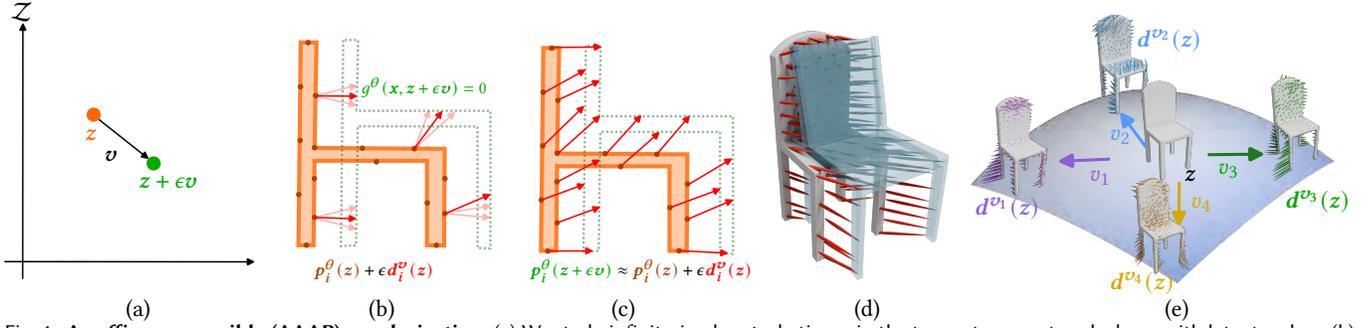

The main goal for learning the shape generative model is to enforce a deformation loss between adjacent synthetic shapes. Implicit deformation regularizations exist in the literature. One notable example is the Killing formulation~\cite{Kraevoy:2008:NRC,SlavchevaBCI17} for as-rigid-as-possible (ARAP) deformation, in which the Jacobian of the implicit field is askew-symmetric. However, this approach does not apply to the affine setting, in which we do not have any constraint on the Jacobian. To address this issue, our approach turns an implicit generation model into an explicit one locally by computing dense correspondences between adjacent synthetic surfaces (see Section~\ref{Subsubsec:AAAP:Deformation}). These correspondences give us flexibility in enforcing the  piece-wise affine assumption, and we define a regularization loss using a robust AAAP model to learn the shape generator (see Section~\ref{Subsubsec:Shape:Generator:Learning}). In addition, we present a lightweight test-time optimization strategy for test shapes in Section~\ref{Subsubsec:Shape:Generator:Finetuning}.

\subsubsection{AAAP deformation induced correspondences}
\label{Subsubsec:AAAP:Deformation}

We first explain how to compute the correspondences between adjacent surfaces of an implicit generator model $g^{\theta}(\bs{x},\bs{z}): \R^3 \times \set{Z}\rightarrow \R$ that outputs the signed distance function (SDF) value at $\bs{x}$ of the underlying 3D shape with the latent code $\bs{z}\in \set{Z}\cong \R^q$ ($q=256$ in our experiments).  We focus on an AAAP model for man-made shapes. 

To this end, we first discretize $g^{\theta}(\bs{x},\bs{z}) = 0$ using a mesh with $n \approx 2000$ vertices $\bs{p}_i^{\theta}(\bs{z}), 1\leq i \leq n$, e.g., via Marching Cube~\cite{Lorensen:1987:MC}. To define the AAAP deformation model, our goal is to determine the corresponding location $\bs{p}_i^{\theta}(\bs{z}+\epsilon\bs{v})$ on the neighboring implicit surface $g^{\theta}(\bs{x},\bs{z}+\epsilon\bs{v}) = 0$, where $\bs{v}$ is the direction of the perturbation and $\epsilon=10^{-3}$ is an infinitesimal value. We approximate $\bs{p}_i^{\theta} (\bs{z}+\epsilon\bs{v})$ via the first-order Taylor expansion $\bs{p}_i^{\theta}(\bs{z}+\epsilon\bs{v})\approx \bs{p}_i^{\theta}(\bs{z}) + \epsilon\bs{d}_i^{\bs{v}}(\bs{z})$.

By computing the derivative of the implicit function $g^{\theta}(\bs{p}_i^{\theta} (\bs{z}+\epsilon\bs{v}),\bs{z}+\epsilon\bs{v}) = 0$ with respect to $\bs{v}$ using the chain rule, we arrive at the following linear constraint of $\bs{d}_i^{\bs{v}}(\bs{z})$: 
\begin{equation}
\frac{\partial g^{\theta}}{\partial \bs{x}}\big(\bs{p}_i^{\theta}(\bs{z}),\bs{z}\big)^T\bs{d}_i^{\bs{v}}(\bs{z}) +  \frac{\partial g^{\theta}}{\partial \bs{z}}\big(\bs{p}_i^{\theta}(\bs{z}),\bs{z}\big)^T\bs{v} = 0.
\label{Eq:Implicit:Cons}    
\end{equation}

As shown in Figure~\ref{Figure:AAAP:Detail}(b), the technical challenge is that the implicit surface representation only provides one constraint on the 3D displacement vector $\bs{d}_i^{\bs{v}}(\bs{z})$. To address the uniqueness problem, we solve a linearly constrained optimization problem~\cite{DBLP:journals/cgf/TaoSB16,Yang:2023:ICCV} to find $\bs{d}_i^{\bs{v}}(\bs{z})$ jointly.

Specifically, we associate each vertex $\bs{p}_i^{\theta}(\bs{z})$ with a local transformation $I_3 + A_i \in \R^{3\times 3}$, where $A_i$ encodes the deviation of this transformation from the identity transformation. For the sake of optimization, we reparameterize $A_i = s_i I_3 + \bs{c}_i\times+ R(\bs{a}_i)$, where $s_i I_3$ and $\bs{c}_i\times$ define scaling and rotation components, respectively; $R(\bs{a}_i)$ denotes the remaining component in the matrix space that is orthogonal to conformal matrices, and $\bs{a}_i \in \R^5$ is the coefficient vector under an orthonormal basis of this space (see Appendix ~\ref{Section:Expression} in the supp. material). A nice property of this parametrization is that it is easy to regularize $A_i$ using quadratic objective terms. 

Let $\set{N}_i$ denote the neighbors of the vertex $i$ and itself. Let vectors $\bs{p}^{\theta}(\bs{z})\in \R^{3n}$ and $\bs{d}^{\bs{v}}(\bs{z})\in \R^{3n}$ concatenate $\bs{p}_i^{\theta}(\bs{z})$ and $\bs{d}_i^{\bs{v}}(\bs{z})$, respectively. We model the deformation energy between them as 
\begin{align}
e\big(\bs{p}^{\theta}(\bs{z}),\bs{d}^{\bs{v}}(\bs{z})\big):=& \min\limits_{\{A_i\}} \sum\limits_{i=1}^{n}\Big(\sum\limits_{j \in \set{N}_i} \|A_i\big(\bs{p}_i^{\theta}(\bs{z})-\bs{p}_{j}^{\theta}(\bs{z})\big) \nonumber \\
&- \big(\bs{d}_i^{\bs{v}}(\bs{z})-\bs{d}_{j}^{\bs{v}}(\bs{z})\big)\|^2 
   + \big(\mu_r s_i^2 + \mu_s\|\bs{a}_i\|^2 \big)\Big)
\label{Eq:E:Obj}
\end{align}
where $\mu_r = 1$ and $\mu_s = 1$ are regularization parameters. Our experiments show that the regularization effects remain similar in a large range of $\mu_r$ and $\mu_s$. On the other hand, regularization is important since otherwise $A_i$ maybe degenerate, e.g., on flat regions.

\begin{figure}[b]
\includegraphics[height=0.38\linewidth]{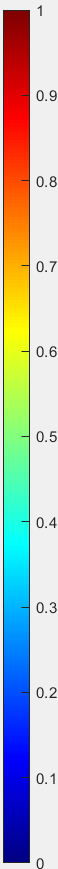}
\includegraphics[height=0.38\linewidth]{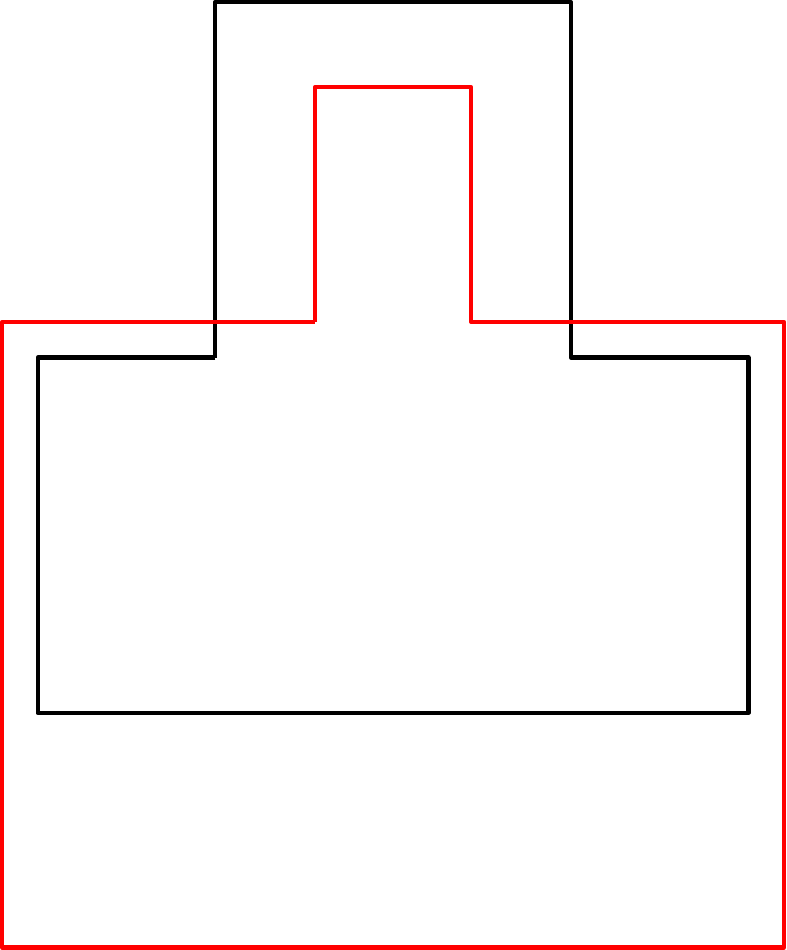}
\includegraphics[height=0.38\linewidth]{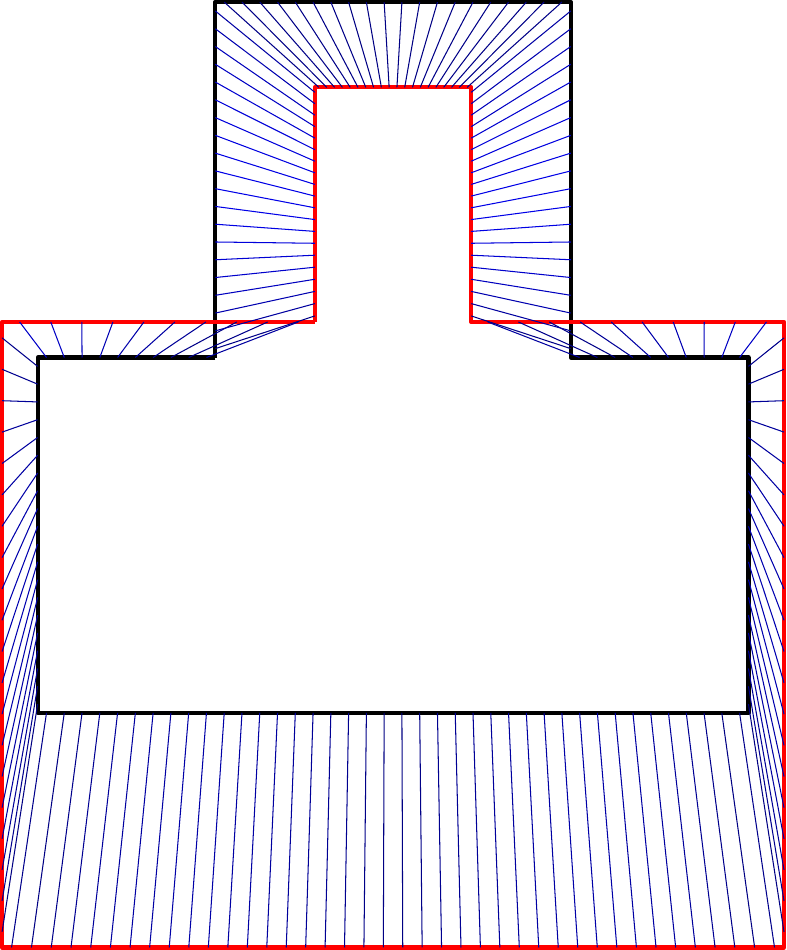}
\includegraphics[height=0.38\linewidth]{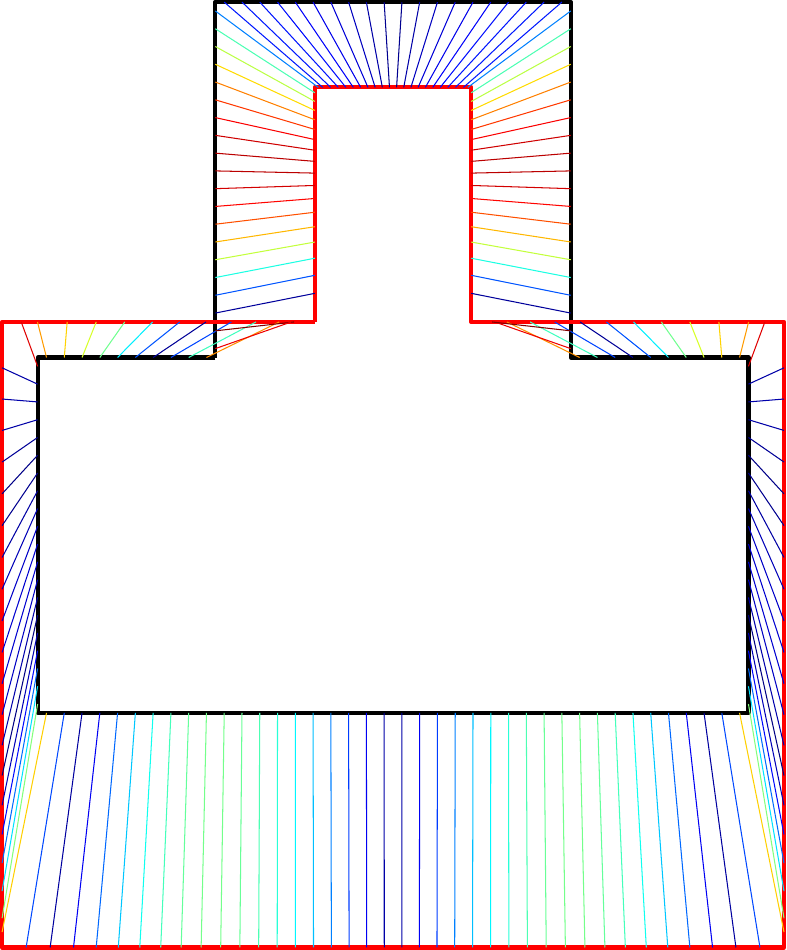}
\caption{Effectiveness of the AAAP formulation for identifying correspondences between two implicit shapes that have two parts, each of which undergoes an affine transformation.  Correspondence errors are color-coded. (Left) The source shape is colored in black. The target shape is colored in red. (Middle) Deform the source shape to align with the target shape under the AAAP model, resulting in accurate correspondences. (Right) Correspondences drift under the ACAP model, i.e., $\bs{a}_i = 0$. Errors are color coded.}
\label{Figure:AAAP:Corres}    
\end{figure}

We introduce $\bs{y}\in \R^{9n}$ where $\bs{y}_{i} = (s_i;\bs{c}_i;\bs{a}_i)$ collects the affine transformation parameters for each $A_i$ at $\bs{p}_i^{\theta}(\bs{z})$. It is clear that $e\big(\bs{p}^{\theta}(\bs{z}),\bs{d}^{\bs{v}}(\bs{z})\big)$ is quadratic in $\bs{y}$ and $\bs{d}^{\bs{v}}(\bs{z})$. Therefore, the resulting $e\big(\bs{p}^{\theta}(\bs{z}),\bs{d}^{\bs{v}}(\bs{z})\big)$ could be written as
\begin{equation}
e\big(\bs{p}^{\theta}(\bs{z}),\bs{d}^{\bs{v}}(\bs{z})\big) = {\bs{d}^{\bs{v}}(\bs{z})}^T L^{\theta}(\bs{z}) \bs{d}^{\bs{v}}(\bs{z}),
\label{Eq:Explicit:Expressions:e}
\end{equation}
and the optimal solution 
\begin{equation}
\quad \bs{y}^{\star} = B^{\theta}(\bs{z})\bs{d}^{\bs{v}}(\bs{z})
\label{Eq:Explicit:Expressions:y}
\end{equation}
where both $L^{\theta}(\bs{z})\in \R^{3n\times 3n}$ and $B^{\theta}(\bs{z})\in \R^{9n\times 3n}$ are sparse matrices (see Appendix~\ref{Section:Expression}  in the supp. material).

With this setup, we can then minimize the quadratic energy in Eq.~(\ref{Eq:Explicit:Expressions:e}) with respect to the linear constraints in Eq.~(\ref{Eq:Implicit:Cons}). This leads to the optimal displacement vector  
\begin{align}
\bs{d}^{\bs{v}}(\bs{z})= M^{\theta}(\bs{z})\bs{v}. \label{Eq:d:Explicit:Expression}
\end{align}
where $M^{\theta}(\bs{z})\in R^{3n\times q}$ can be computed efficiently by LU-factorization of a sparse matrix (see Appendix~\ref{Section:Expression}  in the supp. material).

Figure~\ref{Figure:AAAP:Corres} shows the effects of this AAAP deformation model to find correspondences between shapes that exhibit large piece-wise affine deformations. Note that even if the term residuals of a piece-wise affine deformation exhibit a heavy-tail distribution and should be modeled using a robust norm, the resulting correspondences are accurate under the $L^2$ norm. However, we will see immediately that using a robust norm to formulate the deformation loss on the correspondences for learning the shape generator is important.

\subsubsection{Shape Generator Learning}
\label{Subsubsec:Shape:Generator:Learning}

The training objective for learning the implicit shape generator $g^{\theta}$ is given by
\begin{align}
\min\limits_{\theta} &\quad \frac{1}{|\set{S}|}\sum\limits_{S\in \set{S}_{\textup{train}}} l_{\textup{data}}\big(S, g^{\theta}(\cdot,\bs{z}_S)\big) + \lambda_{\KL}l_{\textup{\KL}}\big(\{\bs{z}_S\},\set{N}(\bs{0},I_d)\big) \nonumber \\
& + \lambda_{d} \underset{\bs{z}\sim \set{N}(\bs{0},I_q)}{\E} r(\theta, \bs{z}). 
\label{Eq:Training:Loss}    
\end{align}
The first term $l_{\textup{data}}(\cdot, \cdot)$ in Eq.~(\ref{Eq:Training:Loss}) aligns the generator with the input shapes. Given an input shape $S$, let $\set{P}_S = \{(\bs{p},s)\}$ collect samples $\bs{p}$ and their corresponding SDF values $s$ using the DeepSDF~\cite{ParkFSNL19} strategy. We define the data term as
\begin{align}
l_{\textup{data}}(S,g^{\theta}(\cdot, \bs{z}_S)) =  \sum\limits_{(\bs{p}, s)\in P_S} |g^{\theta}(\bs{p},\bs{z}_S)-s|^2.
\label{Eq:Standard:Data:Loss}    
\end{align}

\begin{figure}[t]
\includegraphics[height=0.57\linewidth]{Figures/Colormap.png}
\includegraphics[height=0.57\linewidth]{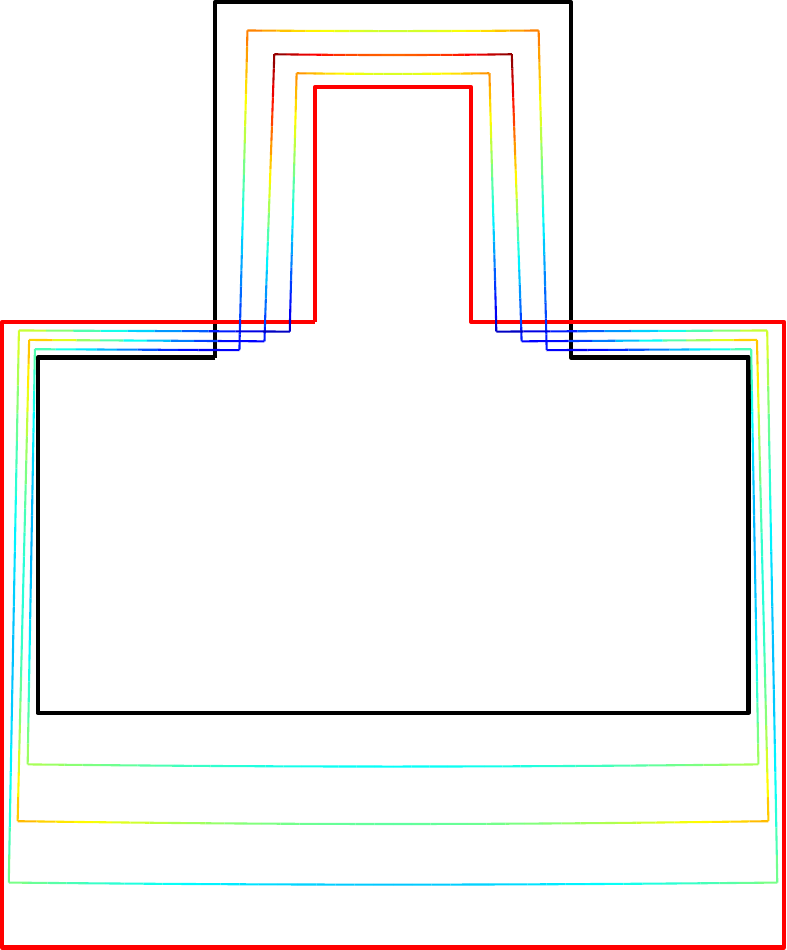}
\includegraphics[height=0.57\linewidth]{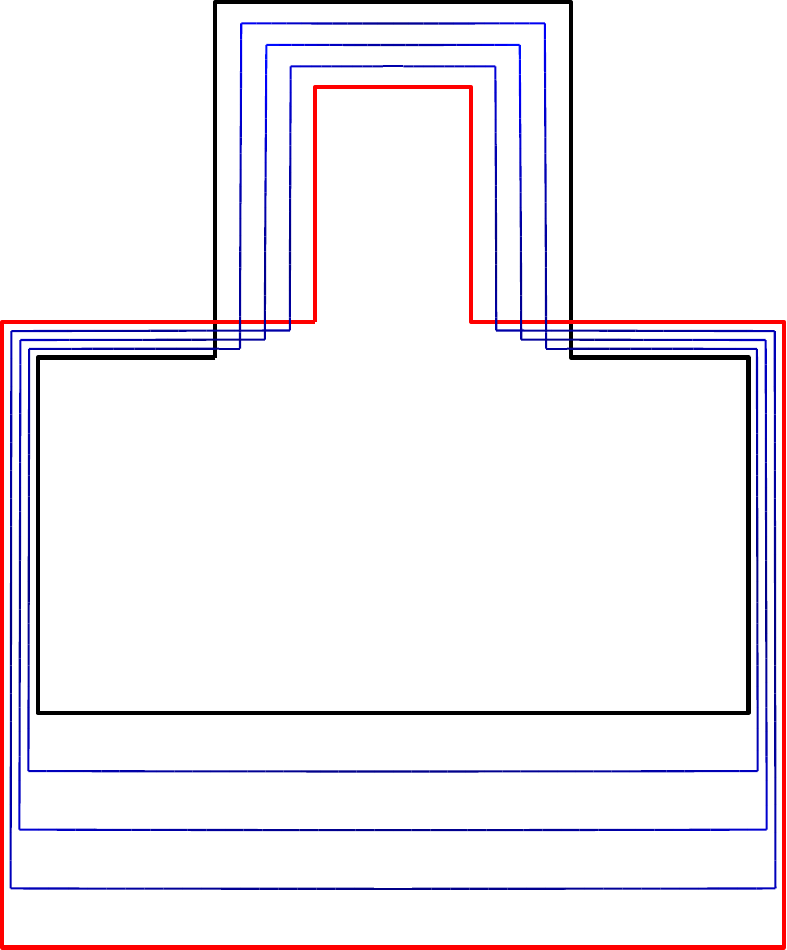}
\caption{Comparison between using different norms to optimize the interpolation between two implicit shapes in Figure~\ref{Figure:AAAP:Corres}. Deviations from the underlying piece-wise affine interpolation are color coded. (Left) Interpolation using the $L^2$ norm deviates from piece-wise affine. (Right) Interpolation using the robust norm is piecewise affine. }
\label{Figure:Robust:Norm}    
\end{figure}

\begin{figure}[t]

\begin{overpic}[width=1.0\columnwidth]{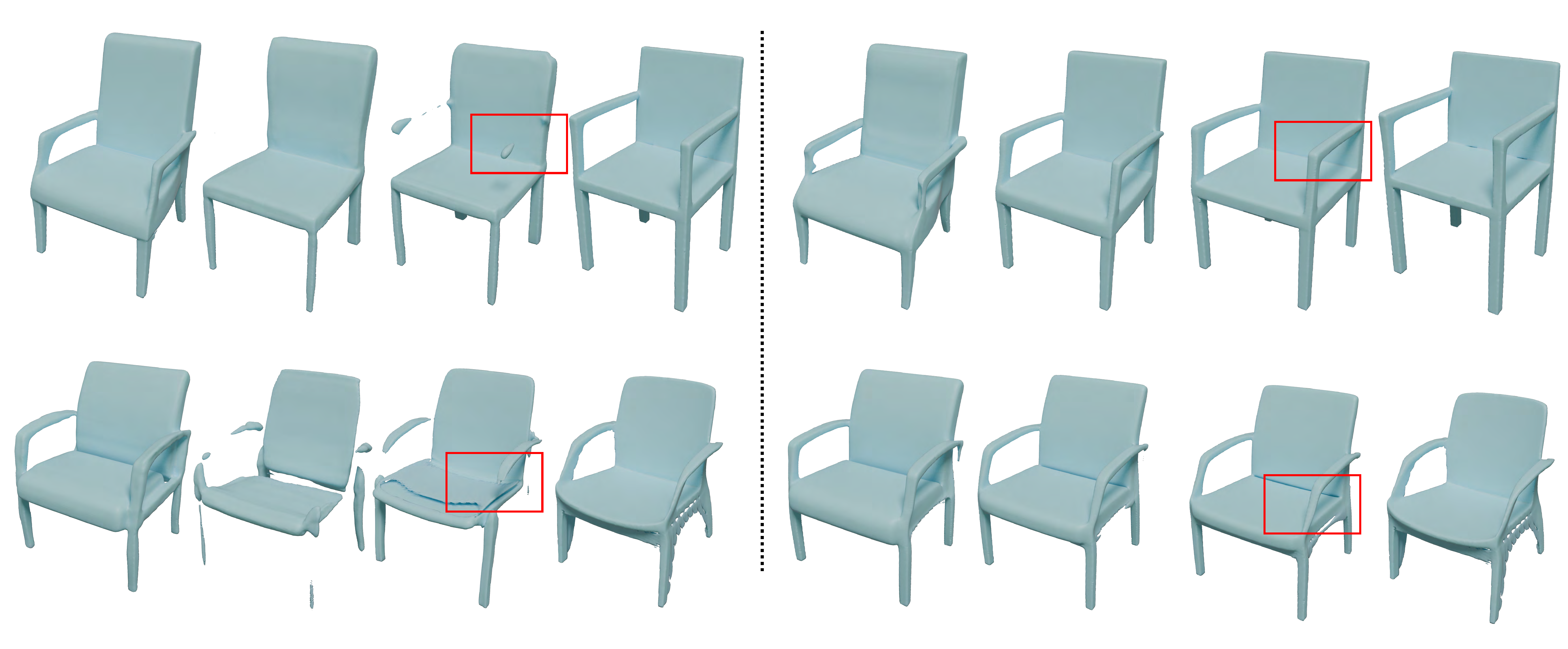}
\put(60, -1){with AAAP reg.}
\put(12, -1){w/o AAAP reg.}

\end{overpic}
\caption{Shape interpolations from the shape generator. (Left)Without AAAP regularization, the intermediate shapes may not preserve part structures. (Right) With AAAP regularization, the part structures are preserved.}
\label{Figure:AAAP:Regularization:Examples}
\end{figure}

The second term $l_{\textup{\KL}}(\cdot,\cdot)$ aligns the empirical latent distribution defined by the training shapes and the prior Gaussian distribution under the KL divergence measure. This enables us to define regularizations on latent codes sampled from the Gaussian distribution.

The third term $r(\cdot, \cdot)$, which is a key contribution of GenAnalysis, enforces AAAP deformations between adjacent synthetic shapes. To formulate $r(\cdot, \cdot)$, we use Eq.~(\ref{Eq:d:Explicit:Expression}), which offers an explicit parameterization of $g^{\theta}(\bs{x},\bs{z}') = 0$ for latent codes $\bs{z}'$ in the local neighborhood of $\bs{z}$. 
Let $\{A_i\}$ be the latent transformations decoded from Eq.~(\ref{Eq:Explicit:Expressions:y}). We define the structure-preserving regularization term $r(\theta,\bs{z})$ to enforce that $\bs{d}^{\bs{v}}(\bs{z})$ admits a piece-wise affine structure via
\begin{align}
r(\theta,\bs{z}) & := \int_{\bs{v}\in \set{B}^q}\sum\limits_{i=1}^{n}\sum\limits_{j\in \set{N}_i} r_{ij}^{\alpha}(\bs{z},\bs{v})d\bs{v},
\label{Eq:Geo:Regu} \\
r_{ij}(\bs{z},\bs{v}) &: = \|A_i \big(\bs{p}_i^{\theta}(\bs{z})-\bs{p}_{j}^{\theta}(\bs{z})\big) - \big(\bs{d}_i^{\bs{v}}(\bs{z})-\bs{d}_{j}^{\bs{v}}(\bs{z})\big)\|.\nonumber 
\end{align}
where $\set{B}^q$ is the unit ball in $\R^q$. We set $\alpha = 1$ to promote a heavy-tailed distribution in $r_{jk}^{\alpha}(\bs{z},\bs{v})$, which models piece-wise affine deformations. Figure~\ref{Figure:Robust:Norm} shows that the robust norm is critical in obtaining the underlying piece-wise affine interpolations between two implicit shapes. 

Figure~\ref{Figure:AAAP:Regularization:Examples} shows the effectiveness of this regularization term, which preserves part structures. In particular, it addresses the challenge of preserving thin-structures under implicit representations.

\subsubsection{Test-Time Optimization}
\label{Subsubsec:Shape:Generator:Finetuning}

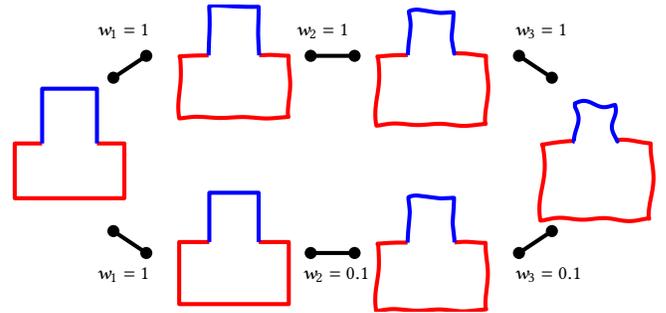
\begin{figure}[t]
  \centering
  
  \captionsetup{aboveskip=2pt, belowskip=-4pt}

  \begin{overpic}[
      width=\linewidth,
      trim=0pt 1pt 0pt 1pt,
      clip
    ]{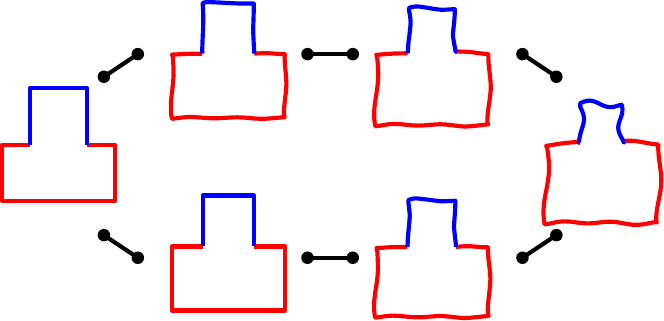}
    \put(13,  5){\footnotesize{$w_{1}=1$}}
    \put(45,  5){\footnotesize{$w_{2}=0.1$}}
    \put(78,  5){\footnotesize{$w_{3}=0.1$}}
    \put(13, 43){\footnotesize{$w_{1}=1$}}
    \put(44, 43){\footnotesize{$w_{2}=1$}}
    \put(78, 43){\footnotesize{$w_{3}=1$}}
  \end{overpic}

  \caption{
    Interpolations between the left and right shapes via two intermediates.
    We minimize AAAP deformations between adjacent shapes.  (Top) Uniform
    weights $(1,1,1)$ cause all deformations to deviate from piece-wise
    affine.  (Bottom) Weights $(1,0.1,0.1)$ emphasize the first pair,
    yielding a near–piecewise‐affine first deformation.
  }
  \Description{Six small meshes arranged in two rows of three.
    The top row shows uniform–weight deformations deviating from piece‐wise
    affine; the bottom row shows first‐pair‐weighted deformations that
    closely match the affine assumption.}
  \label{Figure:VaryingWeightEffect}
\end{figure}

Inspired by the success of test-time optimization~\cite{
DBLP:conf/iclr/WangSLOD21,DBLP:conf/cvpr/0001WDE22,DBLP:conf/icml/NiuW0CZZT22,DBLP:journals/corr/abs-2307-05014}, we present a test-time optimization approach on a collection of test shapes $\set{S}_{\textup{test}}$, so that 1) the generator offers improved reconstructions of these shapes, and 2) the tangent spaces at these shapes contain displacement fields that exhibit piece-wise affine structures suitable for shape analysis. Our key idea to achieve 2) is to introduce a weight in front of $r(\bs{z},\theta)$ to distribute the distortions of piece-wise affine deformations. Specifically, for the shape $S$ of interest, we set $w(\bs{z}_S) = 1$ and use a small weight $w(\bs{z})$ when the distance between $\bs{z}$ and $\bs{z}_S$ becomes large. Figure~\ref{Figure:VaryingWeightEffect} shows the effects of using this weighting scheme. When the deformation between two shapes has non piece-wise affine components, we can enforce that the deformation between the source shape and its adjacent interpolation is piece-wise affine by using a large weight. 

Now, let us describe the test-time optimization loss. Let $\bs{z}_{S}^{0}$ be the latent code of $S\in \set{S}_{\textup{test}}$ under the pre-trained generator. Test-time optimization amounts to solve the following optimization problem:
\begin{align}
\min\limits_{\theta} & \frac{1}{|\set{S}_{\textup{test}}|}\sum\limits_{S\in \set{S}_{\textup{test}}}\Big(l_{\textup{data}}\big(S,g^{\theta}(\cdot, \bs{z}_S^0)\big)\nonumber \\
& + \mu \int_{\bs{z}\in \set{B}^q(\bs{z}_S^{0},c_1)}\exp(-\frac{\|\bs{z}-\bs{z}_{S}^0\|}{2c_2^2})r(\theta,\bs{z}) d\bs{z}\Big)
\end{align}
where $\set{B}^q(\bs{z}_S^{0},c_1)$ is the ball centered at $\bs{z}_S^{0}$ with radius $r_1$. We set $c_2 = \frac{1}{3}c_1$ and $c_1$ is the median of the shortest distance between the latent code of each test shape and the training shapes. In addition, $\mu = 10^{-1}$.

\subsection{Shape Variation Analysis}

The second stage of GenAnalysis analyzes the variations of each shape derived from the learned shape generator in the first stage. For each shape $S$, let $\set{P}_{S} = \{\bs{p}_{i}\}$ be $n$ samples on the surface of the corresponding reconstruction $g^{\theta}(\cdot,\bs{z}_S)$. We summarize the analysis result in a distance matrix $D_{\set{P}_S}\in \R^{n\times n}$ among all sample points where a small distance indicates that the two corresponding samples belong to the same underlying part.

\begin{figure}
\includegraphics[width=1.0\columnwidth]{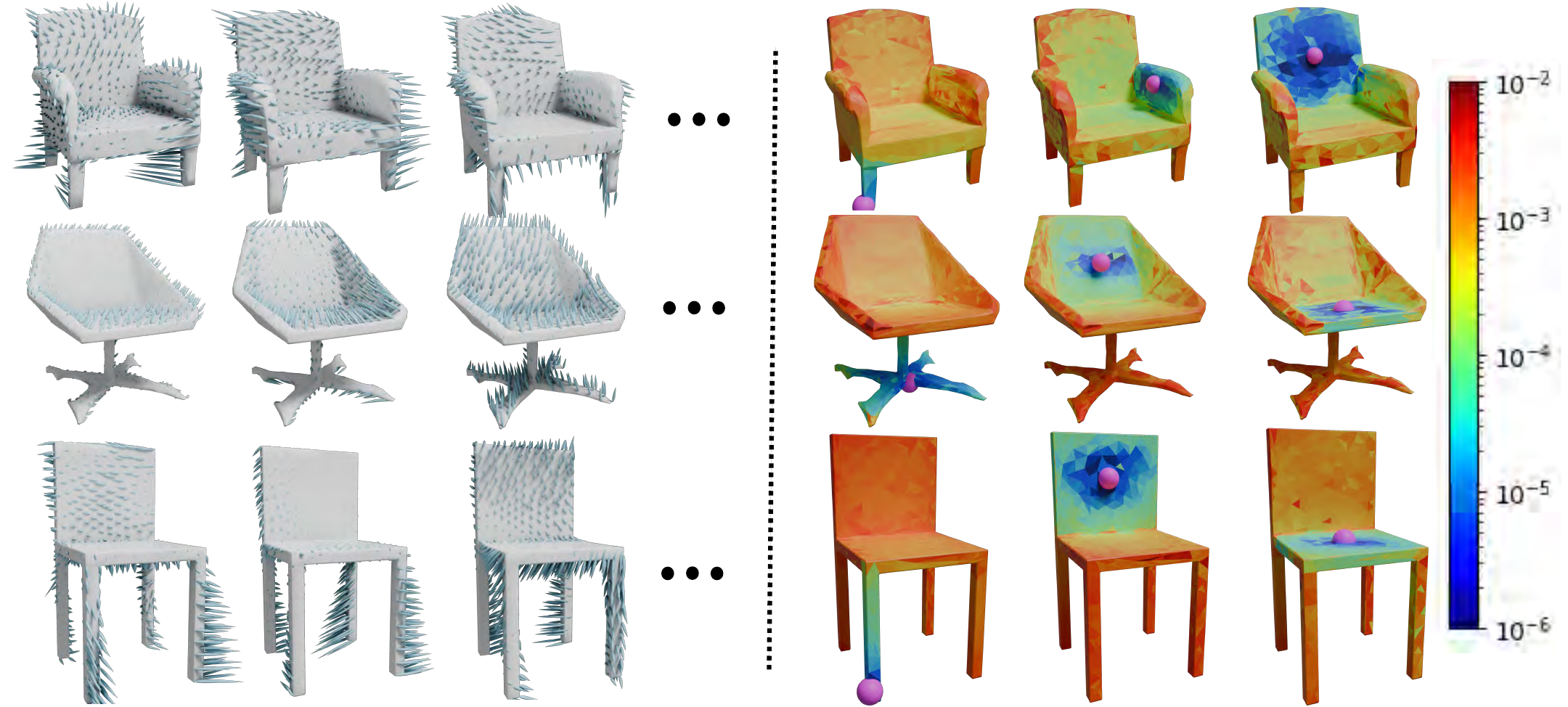}
\caption{(Left) Leading vector fields of example shapes. (Right) Visualizations of distance functions of samples colored in purple, The distance function shows the vector field is piece-wise part awareness. Samples from the same shape part has have smaller distance under affine fitting than samples from different parts.}
\label{Figure:Vector:Fields}
\end{figure}

We compute $D_{\set{P}_S}$ by analyzing piece-wise affine structures in vector-fields $\bs{u}_l = M^{\theta}(\bs{z}_S)\bs{v}_l,1\leq l \leq L$ of $\set{P}_{S}$ ($L=20$ in our experiments), where $M^{\theta}(\bs{z}_S)$ is introduced in Eq.(\ref{Eq:d:Explicit:Expression}). Motivated by the modal analysis framework described in (\cite{Huang:2009:SMA}), we compute $\bs{v}_l, 1\leq l \leq L$ as the smallest eigenvectors of 
$$
H^{\theta}(\bs{z}_S) = {M^{\theta}(\bs{z}_S)}^T L^{\theta}(\bs{z}_S)M^{\theta}(\bs{z}_S).
$$
The behavior of $\bs{u}_l$ is similar to the spectral embedding characterized by leading eigenvectors of a graph Laplacian in which points of the same cluster stay close to each other in the embedding space. In our context, we observe that $\bs{u}_l$ are vector fields that possess affine deformation structures in parts. 

For each sample $\bs{p}_i$ and each vector field $\bs{u}_l = (\bs{u}_{l1};\cdots;\bs{u}_{ln})$, we fit an affine transformation $A_{li},\bs{b}_{li}$ to $\bs{u}_{lj},j\in \set{N}_i$:
\begin{equation}
A_{li},\bs{b}_{li} = \underset{A,\bs{b}}{\textup{argmin}} \sum\limits_{j\in \set{N}_i}\|A\bs{p}_j + \bs{b} - \bs{u}_{lj}\|^2.    
\end{equation}
For any sample $\bs{p}_j$, the residual $\epsilon_{lij} = \|A_{li}\bs{p}_j + \bs{b}_{li} - \bs{u}_{lj}\|$ reveals whether $\bs{p}_j$ and $\bs{p}_i$ belong to the same underlying part or not. When $\epsilon_{lij}$ is small, $\bs{p}_j$ and $\bs{p}_i$ are probably in the same part. In contrast, they are likely in different parts when $\epsilon_{lij}$ is large. In light of this discussion, we define 
$$
D_{\set{P}_S}(i,j) = \big(\sum\limits_{l=1}^{L} w_l \epsilon_{lij}^2\big)^{\frac{1}{2}}. 
$$
where $w_l = \frac{\lambda_{1}(H^{\theta}(\bs{z}_S))}{\lambda_{l}(H^{\theta}(\bs{z}_{S}))}$, and $\lambda_{l}$ is the eigenvalue that corresponds to $\bs{u}_l$. In other words, vector fields with small deformation energies have larger weights. Figure~\ref{Figure:Vector:Fields} (Left) shows vector fields of an example shape. Figure~\ref{Figure:Vector:Fields} (Right) visualizes distances of $D_{\set{P}}(i,j)$ when fixing $i$ while varying $j$. We can see that the resulting distance field is indeed part-aware.

\subsection{Shape Matching}

\begin{figure}[t]
  \centering
  
  \captionsetup{aboveskip=2pt, belowskip=-4pt}

  \begin{overpic}[
      width=\linewidth,
      trim=0pt 0pt 0pt 0pt,  
      clip
    ]{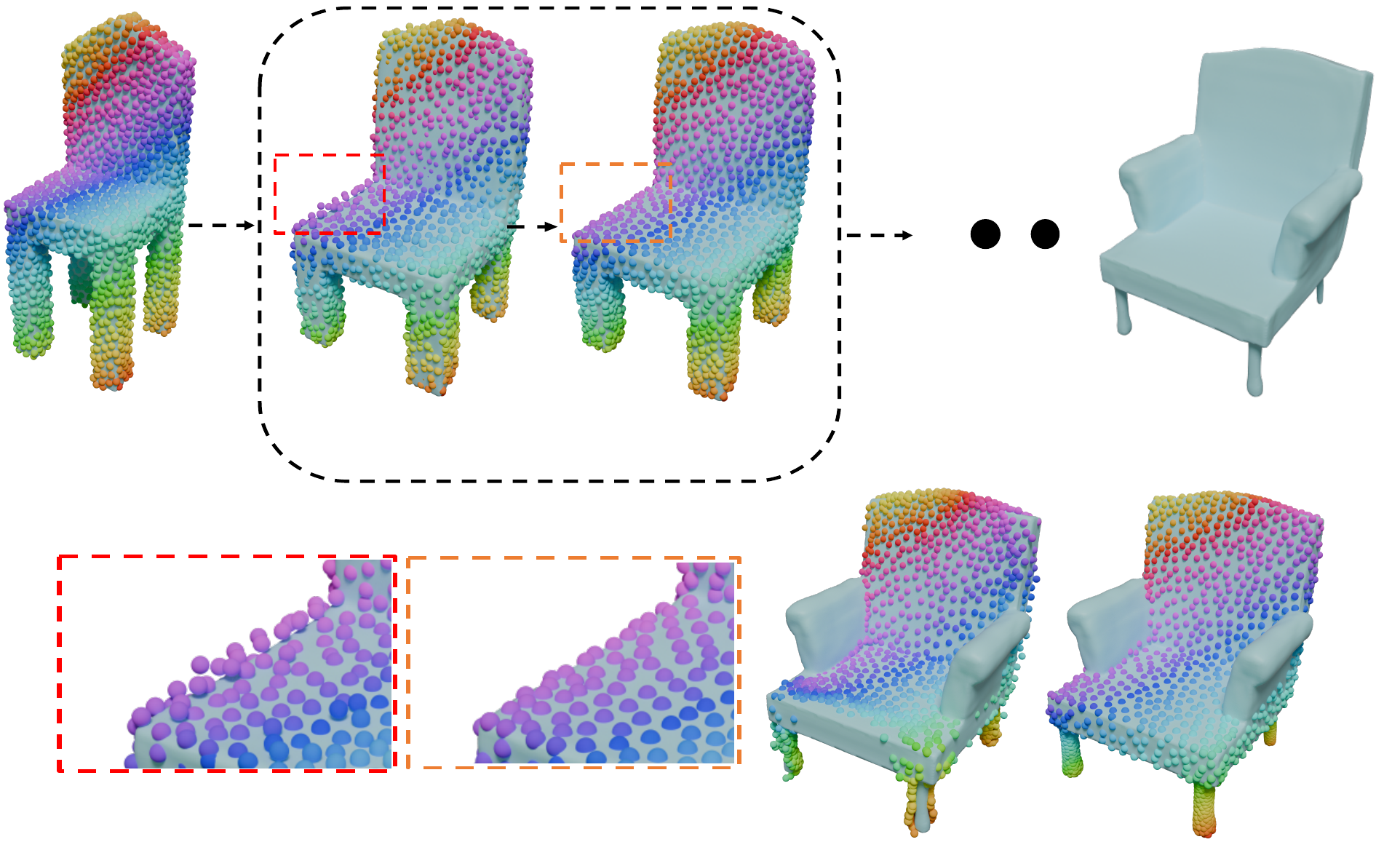}
    \put(0,  29){Source $S$}
    \put(21, 29){Propagate $S^1$}
    \put(43, 29){Project $S^1$}
    \put(82, 29){Target $S'$}
    \put(62, 38){$\times K$}
    \put(52, -1.5){w/o projection}
    \put(76, -1.5){with projection}
  \end{overpic}

  \caption{
    Correspondence computation from the source shape through intermediate
    shapes to the target.  Points share the same colours as they move;
    alternating propagation and projection steps keep them on the surface.
    Without projection, correspondences drift away and fail to lie on the target.
  }
  \Description{A four-panel diagram showing source shape, propagation step,
    projection step, and target shape, with colour‐matched points and labels.}
  \label{fig:propagation-projection}
\end{figure}
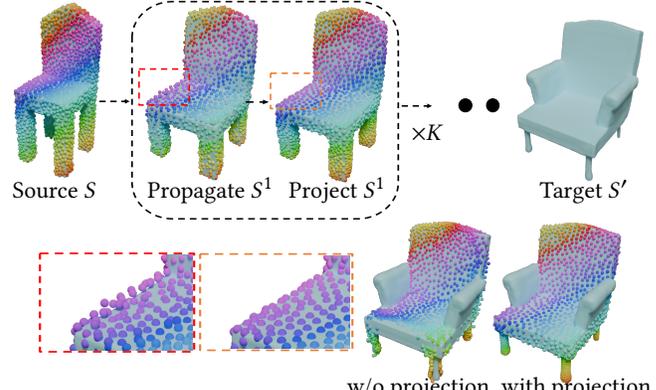

The third stage of GenAnalysis computes inter-shape correspondences between two shapes with latent codes $\bs{z}_{S}$ and $\bs{z}_{S'}$ by mapping samples $\set{P}_{S}$ on $S := \{\bs{x} \mid g^{\theta}(\bs{x}, \bs{z}_{S}) = 0\}$ to the surface $S'$. A naive approach is to apply Eq.~(\ref{Eq:d:Explicit:Expression}) with the latent direction $\bs{v} = \bs{z}_{S} - \bs{z}_{S'}$. However, since Eq.~(\ref{Eq:d:Explicit:Expression}) relies on a linear approximation of the implicit surface constraint, this approach does not result in points lying exactly on the target implicit surface. To address this, we use the generator $g^{\theta}(\cdot, \bs{z})$ to obtain $K = 5$ intermediate shapes $S^{k} := \{\bs{x} \mid g^{\theta}(\bs{x}, \bs{z}^{k}) = 0\}$, where $\bs{z}^{k} = \bs{z}_{S} + \frac{k}{K+1}(\bs{z}_{S'} - \bs{z}_{S})$. We propagate $\set{P}_{S}$ to $S'$ via $S^{k}, 1 \leq k \leq K$. Each propagation step begins by applying Eq.~(\ref{Eq:d:Explicit:Expression}) to compute a displacement vector field for the current sample positions. 

To further mitigate approximation errors at each propagation step, we introduce a projection step to snap the samples onto the surface. Rather than projecting each point in isolation, we solve an optimization problem to ensure consistent projection across all points. This projection step is similar to minimizing the quadratic energy in Eq.(\ref{Eq:Explicit:Expressions:e}) subject to the linear constraints in Eq.(\ref{Eq:Implicit:Cons}). The differences are 1) the quadratic energy is defined using propagated points, and 2) the linear constraints are defined using the next intermediate shape to be projected to.

\begin{figure}[t]
  \centering
  
  \captionsetup{aboveskip=2pt, belowskip=-4pt}

  \includegraphics[
    width=\columnwidth,
    trim=0pt 12pt 0pt 12pt,
    clip
  ]{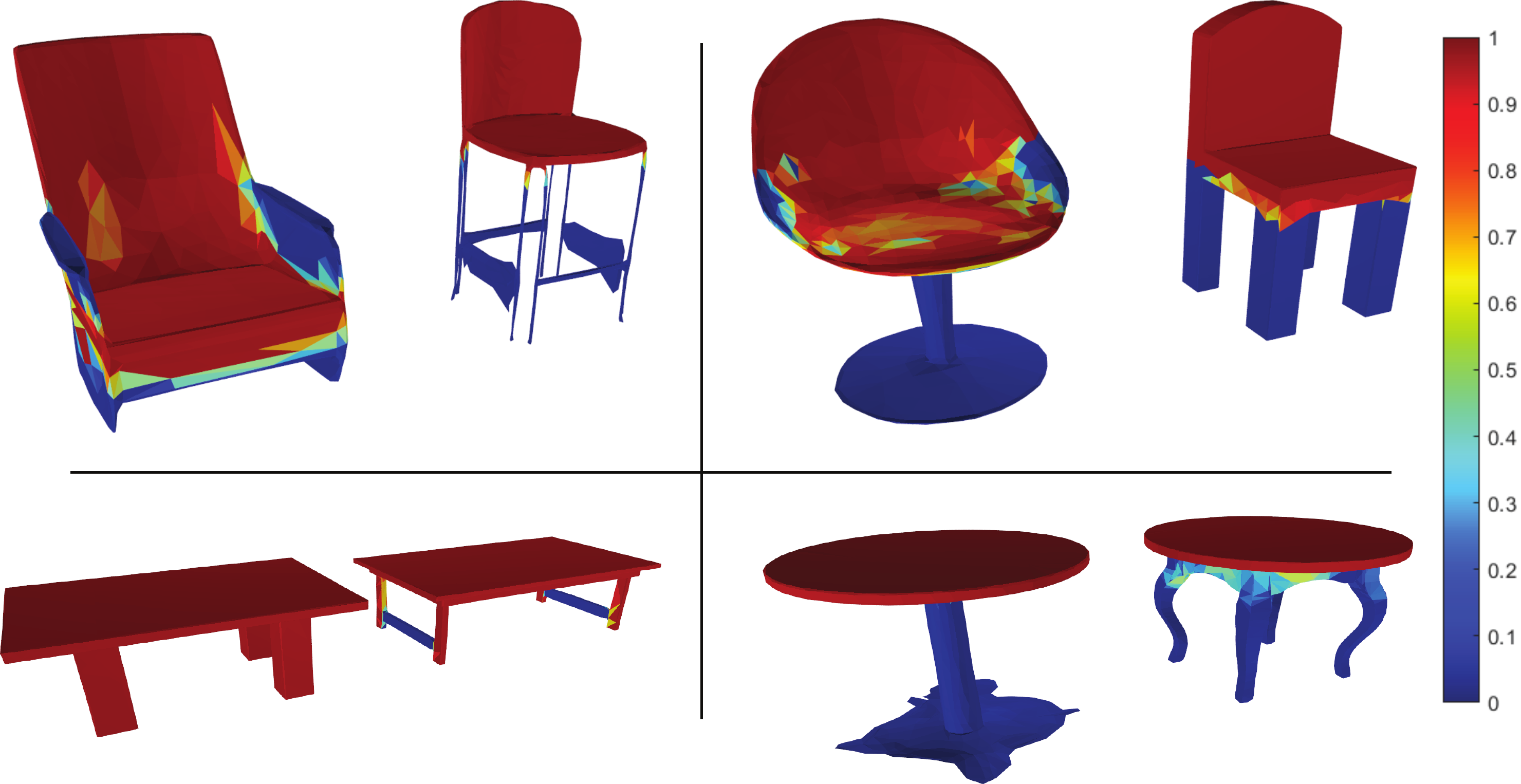}

  \caption{
    Color‑map of similarity weights $w(\boldsymbol{p}_i)$ between the source
    shape (left) and the target shape (right), defined by the local distortions
    $e_i$ of the correspondences computed using our approach.  The similarity
    weights clearly characterize structural similarity across the shapes.
  }
  \Description{Two silhouettes side‑by‑side colored according to
    similarity weights: cooler colours indicate low distortion, warmer colours
    indicate high distortion.}
  \label{Figure:Distortion}
\end{figure}

\begin{figure*}
\begin{overpic}[width=1.0\linewidth]{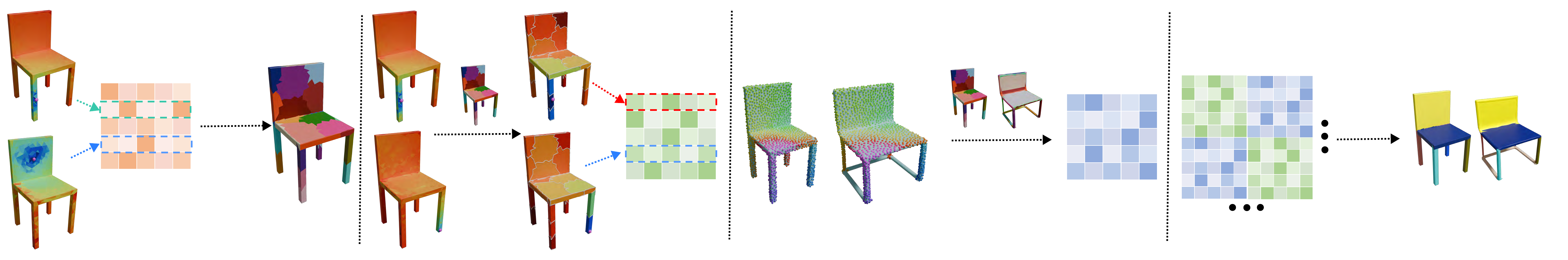}
\put(0, 17){\small (a)}
\put(24, 17){\small (b)}
\put(47, 17){\small (c)}
\put(75, 17){\small (d)}
\put(9, 4){\scriptsize $D_{\set{P}_S}$}
\put(15.5,2){\scriptsize over segement}
\put(24, -1){\scriptsize $D_{\set{P}_S}(i,*)$}
\put(33, -1){\scriptsize $D_{\set{O}_i}(i,*)$}
\put(28,6){\scriptsize aggregate}
\put(61.5,6){\scriptsize aggregate}
\put(15.5,2){\scriptsize over segement}
\put(90,3){\scriptsize consistent segementation}
\put(1, -1){\scriptsize $D_{\set{P}_S}(i,*)$}
\put(42, 4){\scriptsize $W_{ii}$}
\put(70, 4){\scriptsize $W_{ij}$}

\put(52, 2.5){\scriptsize$o(s_i,s_j)$}
\put(76, 12){\scriptsize$W_{00}$}
\put(81, 12){\scriptsize$W_{01}$}
\put(76, 3){\scriptsize$W_{10}$}
\put(81, 3){\scriptsize$W_{11}$}
\put(84.5, 6){\scriptsize graph cut}
\put(12.5, 7.5){\scriptsize graph cut}
\end{overpic}
\caption{\textbf{Overview of our consistent-segmentation algorithm.} (a) We compute over-segments for each shape $S_i$ using its distance matrix $D_{\set{P}_{S_i}}$.   We aggregate (b) affine fitting distance function in each shape and (c) correspondence between each shape pair by over segments.(d) We perform spectral clustering among over-segments of all shapes. Spectral clustering employs an block-wise affinity matrix that encodes each single-shape segmentation cues in its diagonal blocks and correspondences between structurally similar shapes in its off-diagonal blocks. }
\label{Figure:Overview}
\end{figure*}
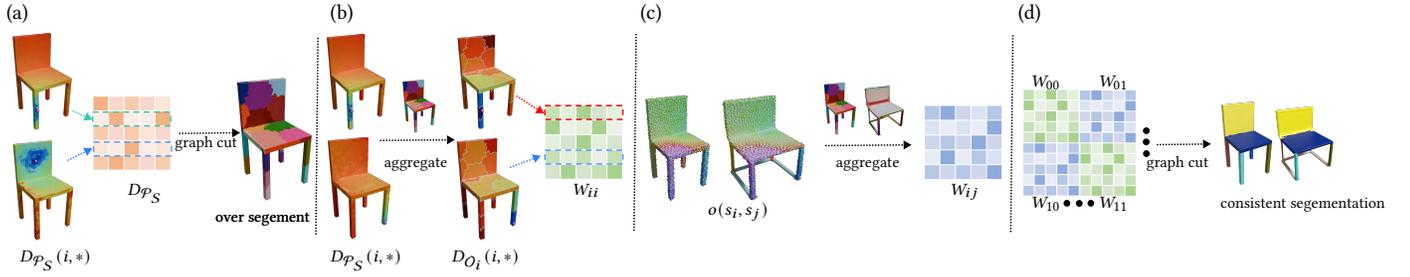

In our implementation, we alternate between one step of propagation and one step of projection. The output of this procedure gives for each point $\bs{p}\in \set{P}_{S}$ its corresponding point on $S'$. When $S$ and $S'$ are structurally similar, such correspondences are meaningful. When $S$ and $S'$ are only partially similar, some of these correspondences are not well defined. We identify such correspondences by calculating the distortion of the neighborhood of each point. In contrast to computing the distortion between $S$ and $S'$ directly, we find that it is more stable to accumulate the distortion during the propagation procedure. For each point $\bs{p}_i$, consider its neighborhood $\bs{p}_j^{k}$ and neighborhood $\bs{p}_j^{k+1}$ at step $k$ and step $k+1$ (after one step of propagation and projection). We first calculate the underlying affine transformation $A_{ik}$ between them by adopting Eq.~(\ref{Eq:E:Obj}):
$$
A_{ik} = \underset{A}{\textup{argmin}} \sum\limits_{j\in \set{N}_i}\|A(\bs{p}_j^{k}-\bs{p}_i^{k})-(\bs{p}_j^{k+1}-\bs{p}_i^{k+1})\|^2 + (\mu_r s^2 + \mu_c\|\bs{a}\|^2).
$$
We then define the distortion as
$$
e_{ik}:= \frac{1}{|\set{N}_i|}\sum\limits_{j\in \set{N}_i}\|A(\bs{p}_j^{k}-\bs{p}_i^{k})-(\bs{p}_j^{k+1}-\bs{p}_i^{k+1})\|.
$$
The similarity weight of the $i$-th sample point is then given by 
\begin{equation}
w(\bs{p}_i) = \exp\big(-e_i^2/2\sigma^2\big), \qquad e_{i} = \sum_{k=1}^{K} e_{ik}
\label{Eq:Corres:Weight}
\end{equation}
where $\sigma$ is the median of $e_i$. Intuitively, a point has high weight if its neighboring patch has small distortions during the propagation procedure. Figure~\ref{Figure:Distortion} illustrates the similarity weights of the source shapes with respect to the target shapes. We can see that the distortions reveal structural similarities and differences.

\subsection{Consistent Segmentation}
\label{Subsec:Cons:Seg}

The last stage of GenAnalysis performs consistent segmentation among a collection of test shapes $\set{S}_{\textup{test}} $ by integrating single-shape segmentation cues formulated in stage II using correspondences obtained in stage III. For computational efficiency concern, we compute $m = 60$ over-segments for each shape (a widely used strategy in image/shape segmentation) and perform consistent segmentation on these over-segments. Figure~\ref{Figure:Overview} illustrates the consistent segmentation procedure. 

\begin{figure}[t]
  \centering
  
  \captionsetup{aboveskip=2pt, belowskip=-4pt}

  \includegraphics[
    width=\columnwidth,
    trim=0pt 12pt 0pt 12pt,  
    clip
  ]{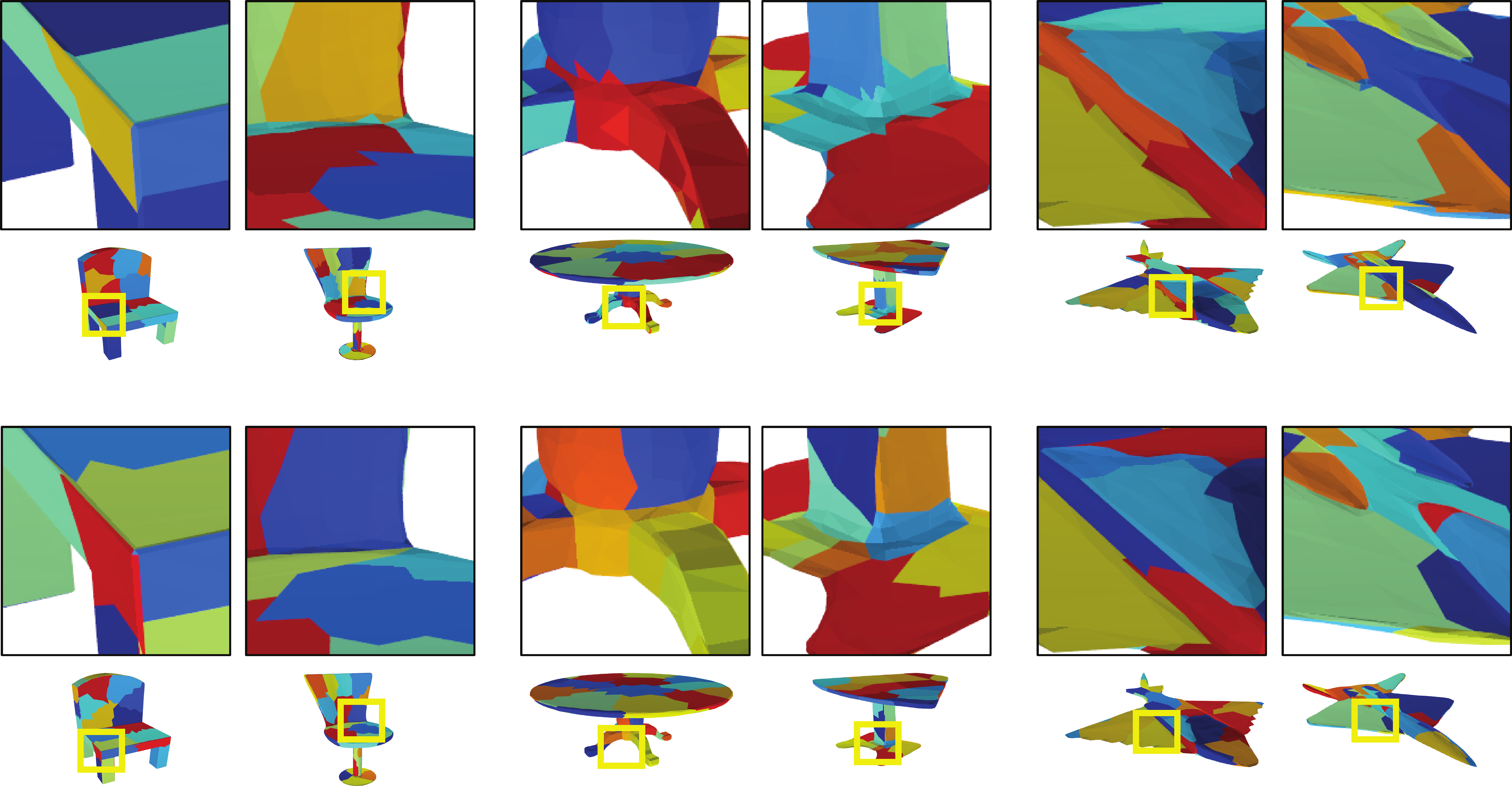}

  \caption{
    Comparison of over‑segmentations from different methods.
    (Top) Segments based on angles between adjacent faces~\cite{Golovinskiy:2008:RC}.
    (Bottom) Segments produced by our approach, which yield more meaningful
    boundaries.
  }
  \Description{Two rows of segmented shapes: the top row shows angular‐based
    over‑segmentation, the bottom row shows our improved segment boundaries.}
  \label{Figure:Over:Segment}
\end{figure}

The over-segments are computed for each shape $S$ in isolation. We feed the distance matrix $D_{\set{P}_S}$ into NormalizedCut~\cite{Shi:2000:NC} to obtain the over-segments $\set{O}_{S}$ for $S$. As shown in Figure~\ref{Figure:Over:Segment}, the resulting over-segments are better than those derived from angles between adjacent faces~\cite{Golovinskiy:2008:RC}.

Consistent segmentation is achieved by performing spectral clustering on an affinity matrix $W \in \R^{(|\set{S}_{\textup{test}} |m)\times (|\set{S}_{\textup{test}} |m)}$, an $|\set{S}_{\textup{test}} |\times |\set{S}_{\textup{test}} |$ block matrix. Each diagonal block $W_{ii}$ encodes the segmentation cue of $S_i\in \set{S}$:
\begin{equation}
W_{ii}(s,s') = \exp\big(-\frac{D_{\set{O}_i}(s,s')^2}{2\overline{\sigma}^2}\big), \qquad 1\leq s, s' \leq m.    
\label{Eq:A:Diagonal:Block}
\end{equation}
where the distance $D_{\set{O}_i}(s,s')$ between over-segments $s$ and $s'$ is given by the mean value of $D_{\set{P}_{S_i}}(i,j)$ where $i\in s$ and $j\in s'$. $\overline{\sigma}$ is the median of $D_{\set{O}_i}(s,s')$ when $s$ and $s'$ are adjacent over-segments.

\begin{figure*}

\begin{overpic}[width=1.0\linewidth]{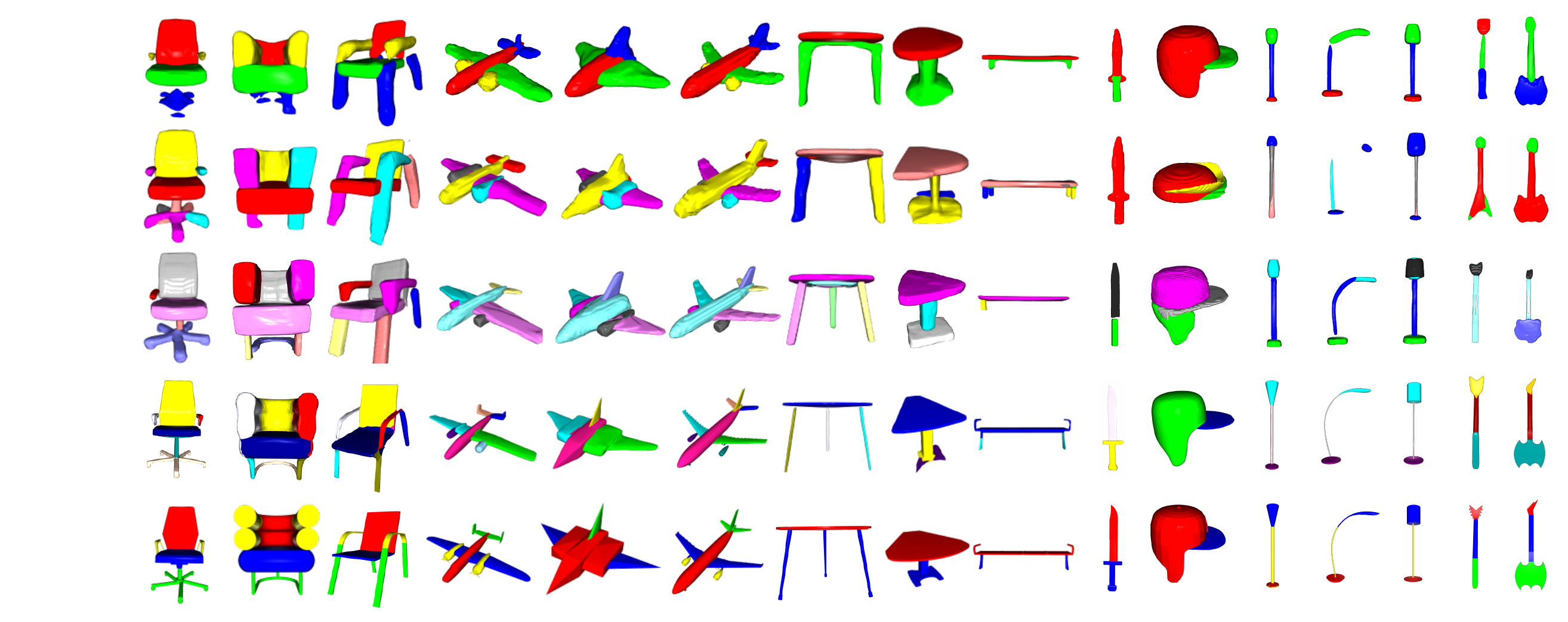}
    \put(0,35){\textbf{BAE-Net}}
    \put(0,28){\textbf{RIM-Net}}
    \put(0,20){\textbf{DAE-Net}}
    \put(1,12){\textbf{Ours}}
    \put(2,4){\textbf{GT}}
    \put(15,-0.5){\small Chair}
    \put(35,-0.5){\small Airplane}
    \put(58,-0.5){\small Table}
    \put(69,-0.5){\small Knife}
    \put(74,-0.5){\small Cap}
    \put(84,-0.5){\small Lamp}
    \put(94,-0.5){\small Guitar}
\end{overpic}
\caption{\textbf{Qualitative evaluation of shape co-segmentation quality on ShapeNet.} We compare co-segmentation results with BAE-Net\cite{Chen2019BAE-Net}, RIM-Net\cite{Niu2022RIM-Net} and DAE-Net\cite{Chen2024DAE-Net}. The colored parts visualize segmentation consistency across different shapes in the same category}    
\label{Figure:Co-Seg}
\end{figure*}

The off-diagonal blocks of $W$ are constructed using a similarity graph whose edges $\set{E}$ connect the nearest neighbors of each testing shape $S\in \set{S}_{\textup{test}}$. In our implementation, we connect 10 most similar shapes, in which the similar score between two shapes is defined as the average of the correspondence weights between them defined in Eq.~(\ref{Eq:Corres:Weight}). We set $W_{ij} = 0, \forall (i,j)\notin \set{E}$. Each non-empty off diagonal block $W_{ij}, (i,j)\in \set{E}$ encodes the correspondences between over-segments of $S_i$ and over-segments of $S_j$. Given two over-segments $s_i\in \set{O}_i$ and $s_j\in \set{O}_j$, we define their affinity score as 
\begin{equation}
W_{ij}(s_i, s_j) = \lambda \frac{|o(s_i,s_j)|}{\max(|s_i|,|s_j|)}\cdot \underset{(\bs{p}_i,\bs{p}_j)\in o(s_i,s_j)}{\textup{mean}}\frac{(w(\bs{p}_i)+w(\bs{p}_j))}{2}
\label{Eq:A:Off:Diagonal:Block}
\end{equation}
where $o(s_i,s_j)$ collects point pairs from $s_i$ and $s_j$ that are in correspondences; the second term in (\ref{Eq:A:Off:Diagonal:Block}) calculates the mean of similarity scores of the correspondences in $o(s_i,s_j)$; $\lambda$ is a hyper-parameter that balances single shape segmentation cues and consistency in segmentations. As the quality of our inter-shape correspondences is high, we set $\lambda = 2$ in our experiments. 

Given $W$, we employ a variant of spectral graph cut to compute clusters of over-segments across all input shapes, which lead to consistent segmentation. Please refer to Appendix~\ref{App:SCS} for details of this spectral clustering procedure. Appendix~\ref{App:CSTable} presents our consistent segmentation algorithm in algorithm block.

\section{Experimental Results}

We demonstrate the effectiveness of GenAnalysis and provide both quantitative and qualitative results. We begin with the experimental setup in Section~\ref{Subsec:Exp:Setup}. We then evaluate the shape matching and shape segmentation results of GenAnalysis in Section~\ref{Subsec:Shape:Corres} and Section~\ref{Subsec:Shape:Seg}, respectively. Section~\ref{Subsec:Ablation:Study} presents an ablation study. 

\subsection{Experimental Setup}
\label{Subsec:Exp:Setup}

\subsubsection{Dataset}
We train and evaluate GenAnalysis and baselines on ShapeNet~\cite{DBLP:journals/corr/ChangFGHHLSSSSX15}. We follow the training and testing split of ShapeNetPart~\cite{10.1145/2980179.2980238}. To evaluate shape matching, we conduct various surrogate tasks including part label transfer and keypoint transfer due to the lack of labeled datasets providing ground truth dense correspondence for direct evaluation. We use labels from ShapeNetPart~\cite{10.1145/2980179.2980238} and KeypointNet~\cite{you2020keypointnet} for the part label transfer and key point transfer tasks, respectively. We evaluate shape matching on three popular categories, i.e., chair, table, and airplane. For shape co-segmentation, we also use labels from ShapeNetPart~\cite{10.1145/2980179.2980238} for evaluation. We report the performance throughout the entire dataset that has 15 categories. The ablation study is performed in the categories of Chair, Table, and Airplane, as do most baseline approaches.

\subsubsection{Baseline approaches} 
For shape matching, we compare our result with the template learning methods, including DIT~\cite{Zheng2021DIT-Net}, DIF~\cite{Deng2021DIF-Net} and Semantic DIF~\cite{Kim2021SemanticDIF-Net} since they achieve the state-of-the-art performance in finding dense correspondences on ShapeNet. We also include AtlasNetV2~\cite{deprelle2019atlasV2} and SIF~\cite{Kyle2019SIF} as additional baselines.

We compare our shape co-segmentation results with BAE-Net~\cite{Chen2019BAE-Net}, RIM-Net~\cite{Niu2022RIM-Net} and DAE-Net~\cite{Chen2024DAE-Net}, which perform unsupervised shape co-segmentation using branched auto-encoding implicit network.  There are some other methods on structure learning that could also perform unsupervised shape co-segmentation. However, they split each input shape into an excessive set of small parts~\cite{Deng2020Cvx-Net,Chen2020BSP-Net}, so they are not compared in our evaluation.

\subsubsection{Implementation Detail} We train our model on a machine with 8 Nvidia Quadro RTX 6000 GPU, and Intel(R) Gold 6230 CPU (40-Core) CPU. We use adam optimizer \cite{DBLP:journals/corr/KingmaB14} and train our model for 4000 epoch using base learning rate of 1e-4. The training takes approximately 3 days to complete. We use the same 8 -layer MLP network as SALD\cite{DBLP:conf/iclr/AtzmonL21} and set $\lambda_{d}$ to 1e-4 and $\lambda_{KL}$ to 1e-5 respectively. Test-time optimization takes 30 minutes on 700 test shapes. Stage II, stage III, and stage IV of GenAnalysis take 1 hour, 4 hours, and 10 minutes, respectively.

\subsection{Analysis of Shape Matching}
\label{Subsec:Shape:Corres}
\setlength\tabcolsep{4pt}

\begin{table}
\caption{\textbf{Label transfer results on three categories of ShapeNetPart.}  We report mean IOU. Higher is better.  Ours-NR means GenAnalysis without AAAP regularization. }
\begin{tabular}{c|c|c|c}

mIOU & Chair& Table& Airplane\\
\hline
  AtlasNetV2~\cite{deprelle2019atlasV2} & 67.1& 59.6& 56.8\\
  SIF~\cite{Kyle2019SIF} & 61.5& 62.7& 54.3\\
  DIT~\cite{Zheng2021DIT-Net} & 79.6& 68.7& 64.4\\
 DIF~\cite{Deng2021DIF-Net} &80.4 & 68.6& 71.9 \\
 SemanticDIF~\cite{Kim2021SemanticDIF-Net} &80.6 & 69.5& 71.8\\\hline
 Ours-NR &80.3 & 70.0& 70.7\\
 Ours &\textbf{82.6} & \textbf{73.0}& \textbf{73.3}\\
 \hline

\end{tabular}

\label{Table:Shape:Labeltransfer} 
\end{table}

\begin{figure}
\begin{overpic}[width=1.0\columnwidth]{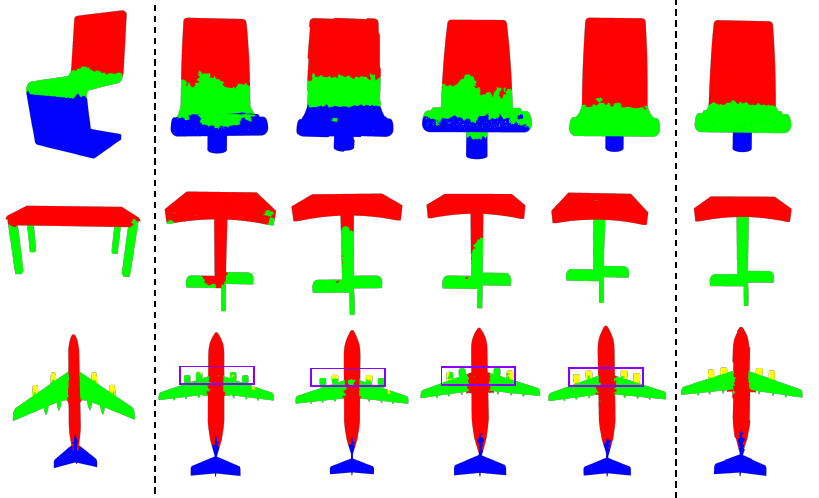}
    \put(4,62){\textbf{Source}}
    \put(23,62){\textbf{DIF}}
    \put(40,62){\textbf{DIT}}
    \put(54,62){\textbf{S-DIF}}
    \put(70,62){\textbf{Ours}}
    \put(88,62){\textbf{GT}}
\end{overpic}
\caption{\textbf{Label transfer results on ShapeNet.} We transfer ShapeNetPart~\cite{10.1145/2980179.2980238} labels from source shapes to target shapes and compare our result with DIF~\cite{Deng2021DIF-Net}, DIT~\cite{Zheng2021DIT-Net} and Semantic DIF~\cite{Kim2021SemanticDIF-Net}.}
\label{Figure:Shape:label_transfer}
\end{figure}

\subsubsection{Part label transfer}
We transfer the part segment labels from the source shape to the target shape using our correspondence results. Specifically, for each of the three categories, we select five labeled shapes as the source shape, transfer their labels to target shapes and compare with ground truth part labels to record the best results. This task can be viewed as 5-shot 3D shape segmentation
using 5 samples as training data. Table~\ref{Table:Shape:Labeltransfer} presents quantitative results on the 5-shot part label transfer task, measured by average per-part IOU. Our method achieves the best performance in all three categories compared to all the baseline methods. Figure~\ref{Figure:Shape:label_transfer} illustrates the qualitative results of part label transfer. AtlasNetV2 and SIF often generate inconsistent correspondences among small regions such as the engine region of airplane and arm region of chair due to the limitation of their shape representations. 
Template-based learning methods~\cite{Deng2021DIF-Net,Zheng2021DIT-Net, Kim2021SemanticDIF-Net} suffer from finding accurate correspondences among structurally distinct and less common shapes due to the limited flexibility of the shared template to represent those shapes. In contrast, GenAnalysis performs much better on those less common shapes with large geometric variations, such as a high-base chair or table. More qualitative visualization results can be found in Figure~\ref{Figure:Shape:label_transfer_supp} in the supp. material.

\begin{table}[h]
\caption{\textbf{Keypoint transfer results on three categories of ShapeNetPart.}  We report the PCK scores with thresholds of 0.01 and 0.02. Higher is better. Ours-NR means GenAnalysis without AAAP regularization. }
\begin{tabular}{c|c|c|c}

PCK& Chair& Table& Airplane\\
\hline
  AtlasNetV2~\cite{deprelle2019atlasV2} & 16.6/37.1& 24.5/45.3& 25.7/42.4\\
  SIF~\cite{Kyle2019SIF} & 20.1/40.7 & 28.6/47.2 & 28.1/46.7 \\
  DIT~\cite{Zheng2021DIT-Net} & 24.6/45.3& 38.9/54.2& 31.7/52.0\\
 DIF~\cite{Deng2021DIF-Net} & 32.9/52.5 & 40.5/61.4 & 36.9/54.7 \\
 SemanticDIF~\cite{Kim2021SemanticDIF-Net} &25.9/44.5 & 26.9/47.9& 20.0/31.1 \\\hline
 Ours-NR &28.9/49.7 & 37.3/57.2& 35.7/61.4\\
 Ours &\textbf{34.9}/\textbf{58.6} & 43.1/\textbf{64.2}& \textbf{40.5}/\textbf{67.8}\\
 \hline

\end{tabular}

\label{Table:Shape:Keypoint}
\end{table}

\begin{figure}[h]
\begin{overpic}[width=1.0\columnwidth]{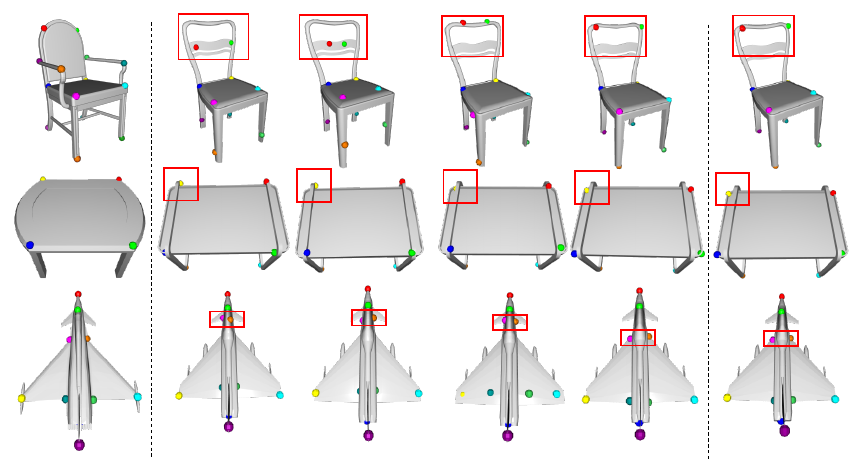}
    \put(4,54){\textbf{Source}}
    \put(22,54){\textbf{DIF}}
    \put(38,54){\textbf{DIT}}
    \put(55,54){\textbf{S-DIF}}
    \put(71,54){\textbf{Ours}}
    \put(88,54){\textbf{GT}}
\end{overpic}
\caption{\textbf{Keypoint transfer results on ShapeNet.} We transfer keypoint labels from source shapes to target shapes and compare our result with DIF~\cite{Deng2021DIF-Net}, DIT~\cite{Zheng2021DIT-Net} and Semantic DIF~\cite{Kim2021SemanticDIF-Net}.}
\label{Figure:Shape:keypoint_transfer}
\end{figure}
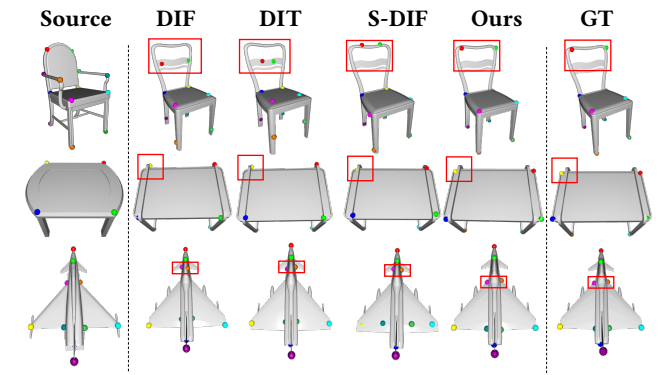

\subsubsection{Keypoint transfer}
Table~\ref{Table:Shape:Keypoint} presents the quantitative result in the keypoint transfer task. We transfer keypoints from the source shape to the target shape using our correspondence results and measure performance by the percentage of correct keypoints (PCK)~\cite{yi2016syncspeccnn}. We compute the geodesic distance between the transferred keypoints and the ground truth points and report the PCK score under a distance threshold of 0.01/0.02. Table~\ref{Table:Shape:Keypoint} shows that our method outperforms baseline approaches. Note that the relevant improvements in keypoint transfer are higher than those in segmentation transfer. This can be understood as the fact that GenAnalysis optimizes point-wise correspondences, while segmentation transfer accuracy does not measure the quality of the correspondences in each segment. Figure~\ref{Figure:Shape:keypoint_transfer} shows the qualitative keypoint transfer results. Our method shows better results in finding the correspondences between structurally distinct shapes, while template-based learning methods \cite{Deng2021DIF-Net,Zheng2021DIT-Net, Kim2021SemanticDIF-Net} fail to find a consistent scale to fit the template and produce inaccurate correspondence results. More qualitative visualization results can also be found in Figure~\ref{Figure:Shape:keypoint_transfer_supp} in the supp. material.

\setlength\tabcolsep{2.5pt}

\begin{table*}[t]
  \centering
  
  \captionsetup{aboveskip=2pt, belowskip=-4pt}
  \caption{
    \textbf{Shape co-segmentation results on ShapeNetPart.}
    We report mean IoU (higher is better). Baselines include
    BAE‑Net~\cite{Chen2019BAE-Net}, RIM‑Net~\cite{Niu2022RIM-Net},
    and DAE‑Net~\cite{Chen2024DAE-Net}.
  }
  \label{Table:Shape:Seg}

  \begin{tabular}{c||c|c|c||c|c|c|c|c|c|c|c|c|c|c|c||c}
    mIoU &
      Chair & Table & Plane & Bag & Cap & Earph. & Guitar & Knife &
      Lamp & Laptop & Motor. & Mug & Pistol & Rocket & Skateb. & Mean\\\hline
    BAE‑Net    & 56.1 & 58.4 & 74.3 & 84.4 & 84.9 & 44.1 & 51.0 & 32.5 &
                 74.7 & 27.1 & 27.5 & 94.4 & 29.0 & 40.9 & 63.3 & 56.2 \\
    RIM‑Net    & 80.2 & 54.9 & 76.0 & \textbf{86.1} & 62.6 & 72.9 & 25.7 &
                 29.5 & 68.7 & 33.2 & 28.5 & 48.6 & 36.2 & 39.5 & 64.9 & 53.6 \\
    DAE‑Net    & 85.5 & 75.5 & 78.0 & 84.4 & 86.3 & 77.2 & 88.4 & 85.8 &
                 73.2 & 95.0 & 48.1 & 94.2 & 74.6 & 38.7 & 68.2 & 76.9 \\ \hline
    GenAnalysis & \textbf{88.4} & \textbf{82.6} & \textbf{79.1} & 86.0 &
                 \textbf{87.3} & \textbf{78.4} & \textbf{92.2} &
                 \textbf{89.8} & \textbf{77.6} & \textbf{97.1} &
                 \textbf{49.6} & \textbf{95.3} & \textbf{74.9} &
                 \textbf{52.7} & \textbf{70.4} & \textbf{80.1} \\
  \end{tabular}
\end{table*}

\subsubsection{Texture transfer}
\begin{figure}[h]
\begin{overpic}[width=1.0\columnwidth]{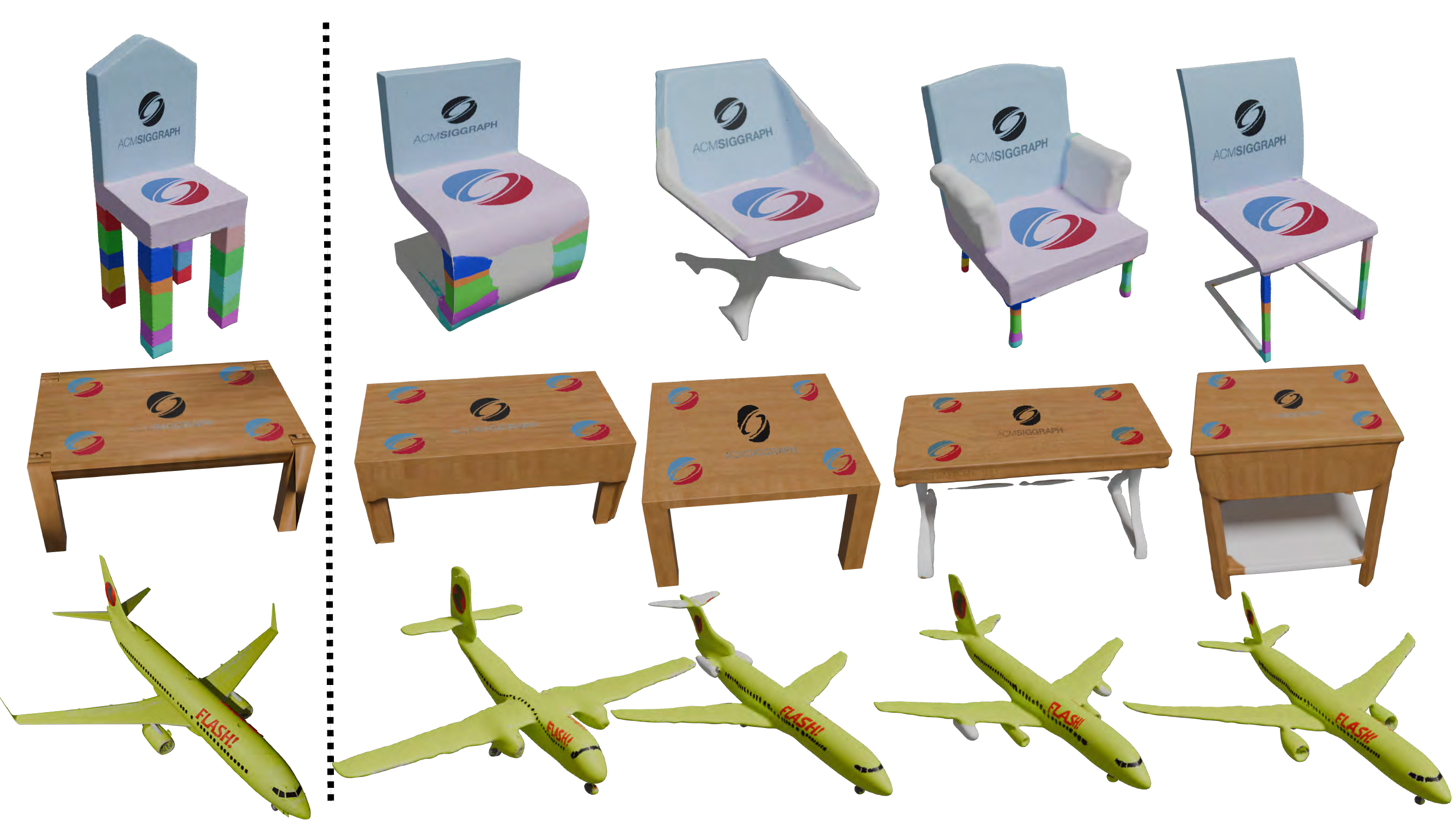}
    \put(4,56){\textbf{Source}}
    \put(53,56){\textbf{Target}}

\end{overpic}
\caption{\textbf{Texture transfer result on ShapeNet.} We transfer texture from source shapes to target shapes with  different structure on ShapeNet objects using coorrespondences generated by GenAnalysis.}
\label{Figure:Shape:texture_transfer}
\end{figure}

We can also transfer mesh attributes such as texture from the source shape to the target shape using the dense correspondence generated by GenAnalysis. Figure ~\ref{Figure:Shape:texture_transfer} shows texture transfer result between shapes in ShapeNet. Our method preserves complex patterns in texture and transfers them to semantically consistent regions.

\subsection{Analysis of Shape Segmentation}
\label{Subsec:Shape:Seg}

We then evaluate the performance of consistent shape segmentation. The same as DAE-Net~\cite{Chen2024DAE-Net}, we quantitatively evaluate the consistent segmentation results by mean IOU in 15 categories of ShapeNetPart.

Table~\ref{Table:Shape:Seg} shows that GenAnalysis achieves the highest accuracy in almost all categories except Bag, in which RIM-Net outperforms GenAnalysis by 0.1\%. In particular, GenAnalysis achieves salient improvements in Chair, Table, Guitar, Knife, Lamp, Laptop, Rocket, and Skateboard, in which part variations exhibit strong piece-wise affine structures. In Laptop, which exhibits perfect piece-wise affine part variations, GenAnalysis reduces the gap to the ground-truth from DAE-Net by 40\% (from 95.0\% to 97.1\%). For other categories including Airplane, Bag, Cap, Earphone, Motorcycle, Mug, and Pistol, in which part variations are more complex, the improvements of GenAnalysis are still noticeable. These results show the robustness of GenAnalysis in different cases. 

We also present qualitative consistent segmentation results in Figure ~\ref{Figure:Co-Seg}. Many baseline methods, especially BAE-Net fail to produce fine-grained segmentations, e.g., four legs of a chair and two back wings of an airplane. These are not reflected in the mean IOU numbers, as these individual segments are grouped into one category in the ground-truth labels.  We also notice that RIM-Net struggles to segment shapes with rare and distinct structures, such as the three-leg table, due to the limited flexibility of the binary tree network. DAE-Net presents inconsistent segmentation results between structurally less similar shapes. In contrast, GenAnalysis produces more consistent and fine-grained segmentation than other baseline methods. Furthermore, our method shows much better reconstruction quality compared to all the baseline methods. In this way, we do not need to perform additional post-processing steps used by other baselines such as projecting labels to ground truth shapes for evaluation. This is because instead of finding a consistent partition of the shape in the template using a bottleneck network, as other baseline methods do, we analyze the piecewise affine variation of the shape to produce consistent segmentation, which does not hurt network reconstruction ability. We show that the piecewise affine transformation assumption fits part variation among a wide range of shape category, which supports the powerful generalization ability of our method to more complex shape collections.

\subsection{Ablation Study}
\label{Subsec:Ablation:Study}

This section presents an ablation study of GenAnalysis. Table~\ref{Table:Ablation:Study} presents quantitative results.  

\subsubsection{Without regularization} As show in Table~\ref{Table:Shape:Keypoint} and Table ~\ref{Table:Shape:Labeltransfer}, the correspondence quality of GenAnalysis in terms of mIOU in the part label transfer task and PCK scores in the keypoint transfer task decreases without the AAAP reguarization loss. This is expected because, without the AAAP regularization, the interpolations from the generator do not recover the underlying deformations between shape pairs. The resulting correspondences easily drift away.

\setlength\tabcolsep{4pt}

\begin{table}[h]
  \centering
  
  \captionsetup{aboveskip=2pt, belowskip=-4pt}

  \caption{
    \textbf{Shape co-segmentation results on ShapeNetPart.}
    We report the mean IoU (higher is better).  
    “GenAnalysis–NR” omits AAAP regularization;  
    “GenAnalysis–No–TanAnal” drops tangent‑space segmentation cues;  
    “GenAnalysis–No–TestTime” removes test‑time optimization;  
    “GenAnalysis–No–Weighting” omits latent‑code weighting at test time.
  }
  \label{Table:Ablation:Study}

  \begin{tabular}{c|c|c|c}
    mIoU & Chair & Table & Airplane \\\hline
    GenAnalysis–NR            & 58.1 & 49.2 & 52.4 \\
    GenAnalysis–No–TanAnal    & 72.3 & 69.2 & 62.4 \\
    GenAnalysis–No–TestTime   & 85.7 & 79.2 & 74.3 \\
    GenAnalysis–No–Weighting  & 87.6 & 81.5 & 77.6 \\\hline
    GenAnalysis               & \textbf{88.4} & \textbf{82.6} & \textbf{79.1} \\
  \end{tabular}
\end{table}

In terms of consistent segmentation, Table ~\ref{Table:Shape:Seg} shows that performance drops in mIOU values are much more significant when the AAAP regularization term is withdrawn, that is, by 30.3\%, 33.4\% and 26.7\% on the Chair, Table, and Airplane.  In addition to having low-quality shape correspondences in this setting, another important factor is that the shape variation segmentation cues become ineffective in this setting.

\subsubsection{Without structural variation cue} Next, we drop the structural variation cue on each input shape and use the edge angle formulation in \cite{Golovinskiy:2008:RC} to generate the over-segments and single shape segmentation cues. In this case, the mIOU values on the Chair, Table, and Airplane drop by 16.1\%, 13.4\%, and 16.7\%, respectively. On the one hand, this shows the effectiveness of the shape variation segmentation cues. However, the decrements are not glaring. This can be understood as the power of consistent shape segmentation using inter-shape correspondences.

\subsubsection{Without test-time optimization} Without the test-time optimization step, the mIOU values drop by a few percentage points. We find that the main issue comes from the fact that the learned shape generator does not offer effective reconstructions of test shapes, although the segmentations on the reconstructed shapes are still good. In this case, performance drops mainly come from projecting the segmentations back to the test shapes. 

\subsubsection{Without latent code weighting} The weighting scheme in test-time optimization improves the mIOU values by 0.8\% to 1.5\% in these three categories. These improvements are consistent, showing its effectiveness.

\begin{figure}
\includegraphics[width=1.0\columnwidth]{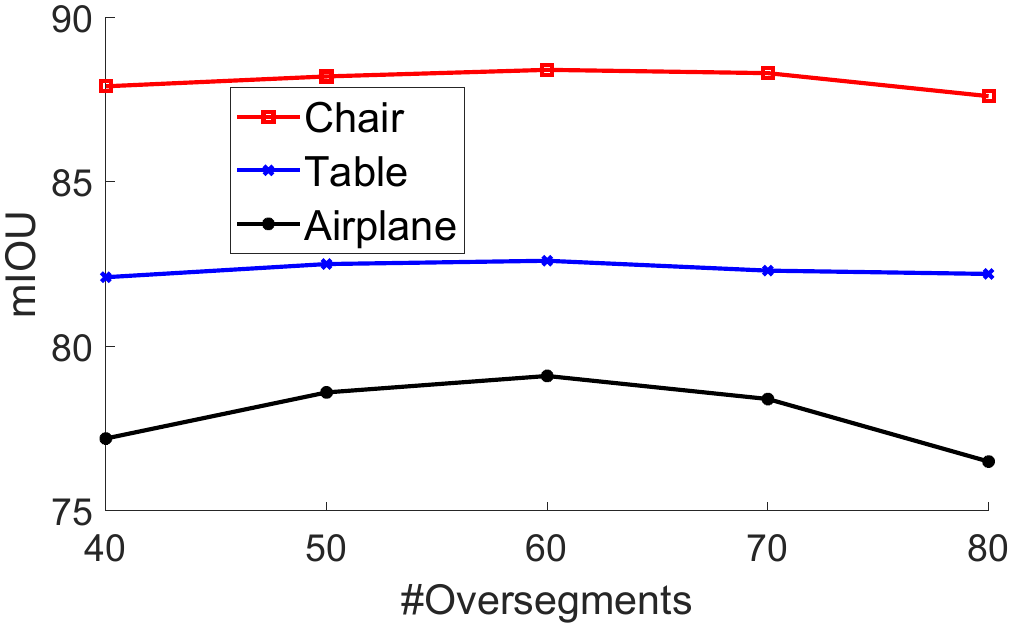}
\caption{\textbf{Mean IOU values when varying the number of over segments.} We show three categories, i.e., Chair, Table, and Airplane.}
\label{Figure:Varying:Number:Oversegments}
\end{figure}

\subsubsection{Varying the number of over-segments $m$} We proceed to analyze the effects when varying the number of over-segments $m$ per shape. Figure~\ref{Figure:Varying:Number:Oversegments} shows the mIOU values of Chair, Table, and Airplane, when varying the number of over-segments $m$. We can see that the values peak around $m = 60$ and drop slightly when $m$ becomes too small or too large. We can understand this behavior as follows. When $m$ is small, the best shape segmentation offered by the over-segments is of low quality. However, the consistent shape segmentation problem, which employs spectral relaxation, is easier to solve, leading to a good approximate solution. In contrast, when $m$ is large, the best shape segmentation offered by the over-segments is of high quality. Yet, it is difficult to obtain the optimal solution, because of the approximation nature of spectral clustering. Nevertheless, the variations in mIOU are not significantly, showing the robustness of GenAnalysis.

\section{Conclusions, Limitations, and Future Work}

In this paper, we have introduced GenAnalysis, a novel framework for performing joint shape analysis by learning implicit shape generators. The key idea is to enforce an as-affine-as-possible (AAAP) deformation prior among neighboring synthetic shapes by establishing correspondences between them. This allows us to establish inter-shape correspondences, to extract structurally similar shapes, and to understand shape variations. We show how to extract single-shape segmentations by recovering piece-wise affine structures from the vector fields in the tangent space of each shape. We also show how to perform consistent shape segmentation by integrating segmentation cues from single-shapes using consistent shape correspondences derived from the shape generator. For both shape matching and shape segmentation, GenAnalysis has achieved state-of-the-art results on ShapeNetPart.

\begin{figure}
\begin{overpic}[width=1.0\columnwidth]{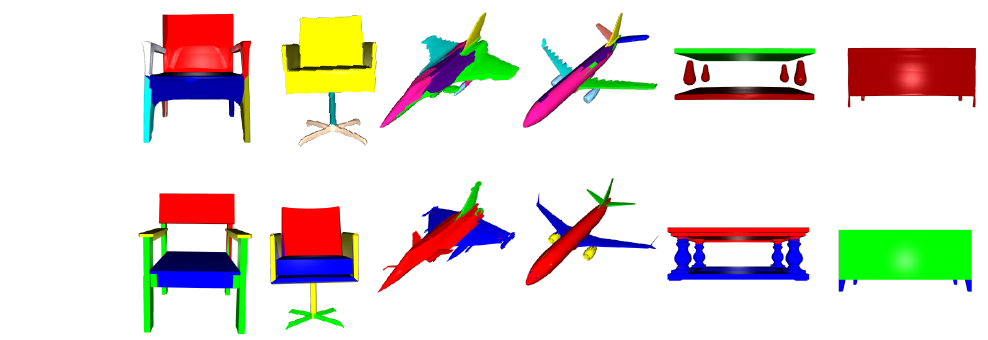}
    \put(2,25){\textbf{Ours}}
    \put(3,7){\textbf{GT}}
\end{overpic}
\caption{\textbf{Unsuccessful co-segmentation results.}}
\label{Figure:Bad:Case}
\end{figure}

The design of GenAnalysis allows efficient test-time optimization that achieves noticeable performance gains. This strength comes from learning an implicit generative model, which can be fine-tuned to fit any test shapes. In contrast, template-based models employed by existing approaches do not possess this property. On the other hand, test-time optimization comes with the cost of optimizing network weights and is computationally more expensive than simply applying learned neural networks during test time.   

Since our method relies on spectral clustering to perform consistent segmentation, for some isolated shapes where shape correspondence quality is poor, our method could perform poorly. See the chair example in Figure ~\ref{Figure:Bad:Case} where the arm of the chair is not distinguished from the back of the chair and the table example. In addition, due to the nature of spectral clustering, we occasionally observe over-segmentations and under-segmentations, as shown in Airplane and Table in Figure ~\ref{Figure:Bad:Case}.

GenAnalysis opens the door to use shape generators to understand shape variations and establish inter-shape correspondences. There are ample opportunities for future research. On the analysis side, we would like to detect clusters of structurally similar shapes and jointly learn a shape generator of each cluster. This can be achieved by decoupling the latent space into a geometry latent code and a structure latent code. The structure latent code can model structural variations, avoiding explicit clustering. Another direction is to explore shape generation. We propose to introduce another latent code to model shape details and enforce AAAP when varying geometry and structure latent codes. 

\begin{acks}
Yuezhi Yang and Qixing Huang were supported by NSF IIS-2047677, NSF IIS-2413161, and gifts from Adobe and Google.
\end{acks}

\bibliographystyle{ACM-Reference-Format}
\bibliography{recons}

\appendix

\section{Expression of the Quadratic Energy}
\label{Section:Expression}

We can express $e\big(\bs{g}^{\theta}(\bs{z}),\bs{d}^{\bs{v}}(\bs{z})\big)$
\begin{align}
:= &\min\limits_{\bs{y}} \ 
\left(
\begin{array}{c}
{\bs{d}^{\bs{v}}(\bs{z})}\\
\bs{y}
\end{array}
\right)^T
\left(
\begin{array}{ccc}
K & E^{\theta}(\bs{z}) \\
{E^{\theta}(\bs{z})}^T &  G^{\theta}(\bs{z})
\end{array}
\right)
\left(
\begin{array}{c}
{\bs{d}^{\bs{v}}(\bs{z})}\\
\bs{y}
\end{array}
\right).
\label{Eq:Quadratic:Energy}
\end{align}
Here $K = \diag(\{|\set{N}_i|\})\otimes I_3$;
$E^{\theta}(\bs{z})$ and $G^{\theta}(\bs{z})$ are $n\times n$ sparse diagonal block matrices, whose expressions are deferred to Appendix~\ref{Expressions:Quadratic:Energy}.

As discussed in Appendix~\ref{Subsec:Proof:Prop:E:Quad:Form}, (\ref{Eq:Quadratic:Energy}) is a quadratic form:
\begin{equation}
e\big(\bs{g}^{\theta}(\bs{z}),\bs{d}^{\bs{v}}(\bs{z})\big) = {\bs{d}^{\bs{v}}(\bs{z})}^T\overline{L}^{\theta}(\bs{z}) \bs{d}^{\bs{v}}(\bs{z}).
\label{Eq:Quad:Obj}
\end{equation}
where
\begin{align}
\overline{L}^{\theta}(\bs{z}) &:= K - E^{\theta}(\bs{z}){G^{\theta}(\bs{z})}^{-1}{E^{\theta}(\bs{z})}^T.
\label{Eq:L:bar:Expression}
\end{align}
Moreover, the optimal transformations are given by
\begin{equation}
\bs{y}^{\star} = - {G^{\theta}(\bs{z})}^{-1}{E^{\theta}(\bs{z})}^T\bs{d}^{\bs{v}}(\bs{z}).   \label{Eq:optimal:y}
\end{equation}

\subsection{Expressions of Matrices in (\ref{Eq:Quadratic:Energy})}
\label{Expressions:Quadratic:Energy}
We can parameterize the elements of $A_i$ as follows:
$$
\textup{vec}(A_i) = \overline{J}\cdot \left(
\begin{array}{c}
s_i \\
\bs{c}_i \\
\bs{a}_i \\
\end{array}
\right) = \overline{J}\bs{y}_i
$$
where
$$
\quad \overline{J}:=\left(
\begin{array}{ccccccccc}
1& 0& 0 & 0& \frac{1}{\sqrt{2}} & -\frac{1}{\sqrt{6}} & 0&0&0\\
0& 0& 0 & 1& 0& 0& 1 &0 & 0\\
0& 0& -1& 0& 0& 0& 0 &1 & 0\\
0& 0& 0& -1& 0& 0& 1 &0 & 0\\
1& 0& 0&  0& 0& \frac{2}{\sqrt{6}}&0 & 0&0\\
0& 1& 0 & 0& 0& 0& 0 &0 & 1\\
0& 0& 1 & 0& 0& 0& 0 &1 & 0\\
0&-1& 0 & 0& 0& 0& 0 &0 & 1\\
1& 0& 0 & 0& -\frac{1}{\sqrt{6}}& -\frac{1}{\sqrt{6}}&0&0&0\\

\end{array}
\right).
$$
Let
$$
R = \left(
\begin{array}{ccc}
\mu_r & 0 & 0 \\
0 & 0 & 0 \\
0 & 0 & \mu_s I_5
\end{array}
\right).
$$
It follows that
\begin{align}
e\big(\bs{g}^{\theta}(\bs{z}),\bs{d}^{\bs{v}}(\bs{z})\big) & = \sum\limits_{i=1}^{n}\Big(\sum\limits_{j\in \set{N}_i}\|((\bs{p}_i^{\theta}(\bs{z})-\bs{p}_j^{\theta}(\bs{z}))\otimes)^T\overline{J}\bs{y}_i\nonumber \\
& \quad -(\bs{d}_i^{\bs{v}}(\bs{z})-\bs{d}_j^{\bs{v}}(\bs{z}))\|^2 + \bs{y}_i^T R\bs{y}_i\Big) \nonumber 
\end{align}
It follows that $K$ is the Laplacian matrix of the graph whose edges are $\{(i,j)|j\in \set{N}_i, 1\leq i \leq n\}$; the diagonal blocks of $G^{\theta}$ are 
$$
G_{ii}^{\theta}(\bs{z}) = \overline{J}^T(\sum\limits_{j\in \set{N}_i}\big(\bs{p}_i^{\theta}(\bs{z})-\bs{p}_j^{\theta}(\bs{z})\big)\big(\bs{p}_i^{\theta}(\bs{z})-\bs{p}_j^{\theta}(\bs{z})\big)^T\otimes I_3)\overline{J} + R;
$$
The $ij$-th block of $E^{\theta}(\bs{z})$ are given by
$$
E_{ij}^{\theta}(\bs{z}) = \left\{
\begin{array}{cc}
-\sum\limits_{j'\in \set{N}_i}(\bs{p}_i^{\theta}(\bs{z})-\bs{p}_{j'}^{\theta}(\bs{z}))^T\otimes I_3& j = i \\
(\bs{p}_i^{\theta}(\bs{z})-\bs{p}_j^{\theta}(\bs{z}))^T\otimes I_3& j\in \set{N}_i\setminus\{i\} \\
0& \textup{otherwise}
\end{array}
\right.\
$$

\subsection{Quadratic Norm Expression of Eq.(\ref{Eq:E:Obj})}
\label{Subsec:Proof:Prop:E:Quad:Form}

Given $\bs{d}^{\bs{v}}(\bs{z})$, the optimal solution to 
$$
\min\limits_{\bs{y}} \ ({\bs{d}^{\bs{v}}(\bs{z})}^T,\bs{y}^T) \left(
\begin{array}{ccc}
K & E^{\theta}(\bs{z}) \\
{E^{\theta}(\bs{z})}^T &  G^{\theta}(\bs{z})
\end{array}
\right)({\bs{d}^{\bs{v}}(\bs{z})}; \bs{y})
$$
is given by
$$
\bs{y} = -{G^{\theta}(\bs{z})}^{-1}
{E^{\theta}(\bs{z})}^T.
$$
Therefore,
\begin{equation}
L^{\theta}(\bs{z}) = K - E^{\theta}{(\bs{z})}^{-1}
{E^{\theta}(\bs{z})}^T 
\label{Eq:A:1}
\end{equation}

\subsection{Correspondence Computation}

We solve the following quadratic minimization problem with linear constraint to find $\bs{d}^{\bs{v}}(\bs{z})$:
\begin{align}
 \bs{d}^{\bs{v}}(\bs{z}):= \lim_{\mu\rightarrow 0}\underset{\bs{d}}{\textup{argmin}} & \quad \bs{d}^T L^{\theta}(\bs{z})\bs{d} + \mu \|\bs{d}\|^2 \nonumber \\
 s.t. & \quad C^{\theta}(\bs{z}) \bs{d} = -F^{\theta}(\bs{z})\bs{v}  
 \label{Eq:d:Q:C:L:P}
\end{align}
where $C^{\theta}(\bs{z}) \bs{d} = -F^{\theta}(\bs{z})\bs{v}$ is the matrix representation of Eq.(\ref{Eq:Implicit:Cons}). $\mu \|\bs{d}\|^2$ avoids degenerate solutions.

It is easy to check the optimal solution to (\ref{Eq:d:Q:C:L:P}) is given by
\begin{align}
\bs{d}^{\bs{v}}(\bs{z})&= M^{\theta}(\bs{z})\bs{v}. \label{Eq:d:Explicit:Expression2}\\
M^{\theta}(\bs{z}) & = - (I_{3n}, 0)\left(
\begin{array}{cc}
L^{\theta}(\bs{z}) & {C^{\theta}(\bs{z})}^T \\
C^{\theta}(\bs{z}) & 0
\end{array}
\right)^{\dagger}F^{\theta}(\bs{z}) 
\label{Eq:Sparse:Inverse}
\end{align}
where $A^{\dagger}$ denotes the pseudo inverse of $A$. Eq.(\ref{Eq:Sparse:Inverse}) only needs the pseudo-inverse of a sparse matrix, which can be computed efficiently by LU pre-factorization.

\begin{algorithm}[t]
\caption{Consistent Segmentation in GenAnalysis}
\label{alg:consistent_segmentation}
\SetKwInOut{Input}{Input}
\SetKwInOut{Output}{Output}

\Input{
  A collection of test shapes $\set{S}_{\textup{test}}=\{S_i\}$, each with a distance matrix $D_{P S_i}$.\\
  Number of over-segments $m$\\
  Similarity graph $\mathcal{E}$ connecting each test shape to its $k$ most similar shapes (e.g., $k=10$).\\
  Hyper-parameter $\lambda$ (e.g., $\lambda=2$).\\
  Median $\sigma$ for distance normalization.\\
}
\Output{
  Consistent segmentation of the test shapes $\set{S}_{\textup{test}}$.
}

\BlankLine
\textbf{Step 1: Over-segmentation}\\
\ForEach{test shape $S_i \in \set{S}_{\textup{test}}$}{
  Apply NormalizedCut \cite{Shi:2000:NC} to $D_{P S_i}$ to obtain $m$ over-segments $O_{S_i}$.
  \[
      O_{S_i} = NormalizedCut(D_{P S_i}).
  \]
}

\BlankLine
\textbf{Step 2: Construct the Block Affinity Matrix $\mathbf{A}$}\\
Initialize a block matrix 
\[
  \mathbf{W} \in \mathbb{R}^{(|\set{S}_{\textup{test}}|\cdot m) \times (|\set{S}_{\textup{test}}|\cdot m)}.
\]

\BlankLine
\textit{(a) Diagonal Blocks}:\\
\ForEach{shape $S_i \in \set{S}_{\textup{test}}$}{
  \ForEach{over-segments $s,\,s' \in O_{S_i}$}{
    Compute 
    \[
      \mathbf{W}_i(s, s') 
      \;=\; \exp\!\Bigl(-\tfrac{D_{O_i}(s, s')^2}{2\,\sigma^2}\Bigr),
    \]
    where
    \[
      D_{O_i}(s, s')
      \;=\;
      \frac{1}{|s| \cdot |s'|}
      \sum_{i \in s} \sum_{j \in s'}
      D_{P S_i}(i, j).
    \]
  }
}

\BlankLine
\textit{(b) Off-Diagonal Blocks}:\\
\ForEach{$(i,j) \in \mathcal{E}$ with $S_i, S_j \in \set{S}_{\textup{test}}$}{
  \ForEach{over-segments $s_i \in O_{S_i}$, $s_j \in O_{S_j}$}{
    Let $o(s_i, s_j)$ be the set of corresponding point pairs. Then
    \[
      \mathbf{W}_{i,j}(s_i, s_j) = \lambda \,\frac{|\,o(s_i, s_j)\,|}{\max(|s_i|, |s_j|)} 
      \;\cdot\; \mathrm{mean}\Bigl(\tfrac{w(\mathbf{p}_i) + w(\mathbf{p}_j)}{2}\Bigr)_{(\mathbf{p}_i,\mathbf{p}_j)\,\in\,o(s_i, s_j)},
    \]
    where $w(\mathbf{p})$ is the correspondence weight for point $\mathbf{p}$.
  }
}
Set $\mathbf{W}_{i,j} = \mathbf{0}$ for all $(i,j) \notin \mathcal{E}.$

\BlankLine
\textbf{Step 3: Spectral Clustering}\\
Perform spectral graph cut on $\mathbf{A}$ to cluster the over-segments across all shapes in $\set{S}_{\textup{test}}$. The resulting clusters yield consistent segment labels for each shape.

\BlankLine
\Return Consistent segmentation of all shapes in $\set{S}_{\textup{test}}$.

\end{algorithm}

\section{Spectral Consistent Segmentation}
\label{App:SCS}

This section presents details of the spectral consistent segmentation procedure used in GenAnalysis. The procedure takes a generalized adjacency matrix $A\in \R^{nm\times nm}$ as input and outputs $K$ partitions $\{1,\cdots, nm\} =\set{I}_1\cdots \set{I}_{M}$. The optimal value for $M$ is determined later. 

Denote $\bs{d} = A\bs{1}$ and $D = \textup{diag}{\bs{d}}$. We first consider the normalized adjacency matrix
$$
\overline{L} = I_{nm} - D^{-\frac{1}{2}}AD^{-\frac{1}{2}}.
$$
With $\bs{u}_l$ and $\overline{\lambda}_l$ as the $l$-th leading eigenvector and eigenvalue of $\overline{L}$, where $1\leq l \leq L$. In our experiments, we choose $L = 10$.

We form an embedding space 
$$
U = (\bs{u}_2,\cdots, \sqrt{\frac{\overline{\lambda}_2}{\overline{\lambda}_L}}\bs{u}_{L})\in \R^{nm\times (L-1)}.
$$
where each row of $U$ provides the coordinate of the corresponding over-segment. We then perform probabilistic K-means with an isotropic covariance matrix to compute the resulting clusters. The initial centers are determined by farthest point sampling (FPS). 

We determined the optimal value $K$ with the maximum value in $\frac{r_{K-1}}{r_{K}}$ where $r_{K}$ is the maximum distance between each point of the cluster center in FPS with $K$ clusters.  

\section{Consistent Segmentation Algorithm Detail}
\label{App:CSTable}
We summarize the consistent segmentation algorithm presented in section \ref{Subsec:Cons:Seg} in algorithm \ref{alg:consistent_segmentation}.

\begin{table}[h]
\centering
\caption{\textbf{Reconstruction error on three categories of ShapeNetPart.} We report chamfer distance at scale 1e-3. Lower is better.}
\label{Table:Reconstruction}
\begin{tabular}{c|c|c|c}
CD (in 1e-3 scale) & Chair & Table & Airplane \\
\hline
BAE~\cite{deprelle2019atlasV2} & 1.316 & 1.915 & 0.593 \\
RIM~\cite{Kyle2019SIF} & 1.320 & 1.699 & 0.630 \\
DAE~\cite{Zheng2021DIT-Net} & 1.244 & 1.579 & 0.328 \\
Ours & \textbf{0.865} & \textbf{1.246} & \textbf{0.235} \\
\hline
\end{tabular}
\end{table}

\section{Reconstruction Quality Evaluation}
We evaluate the representation capability of learned latent space of GenAnalysis by measuring the reconstruction error using the chamfer distance (CD) in Table ~\ref{Table:Reconstruction}. Figure~\ref{Figure:Co-Seg} shows qualitative comparison of the reconstruction quality against baseline methods. Our results indicate a clear improvement over the baseline methods, since our approach avoids model expressivity limitations commonly associated with template-fitting techniques.

\section{Texture Transfer Comparison}
\begin{figure}[h]
\captionsetup{skip=4pt}
\begin{overpic}[width=1.0\columnwidth]{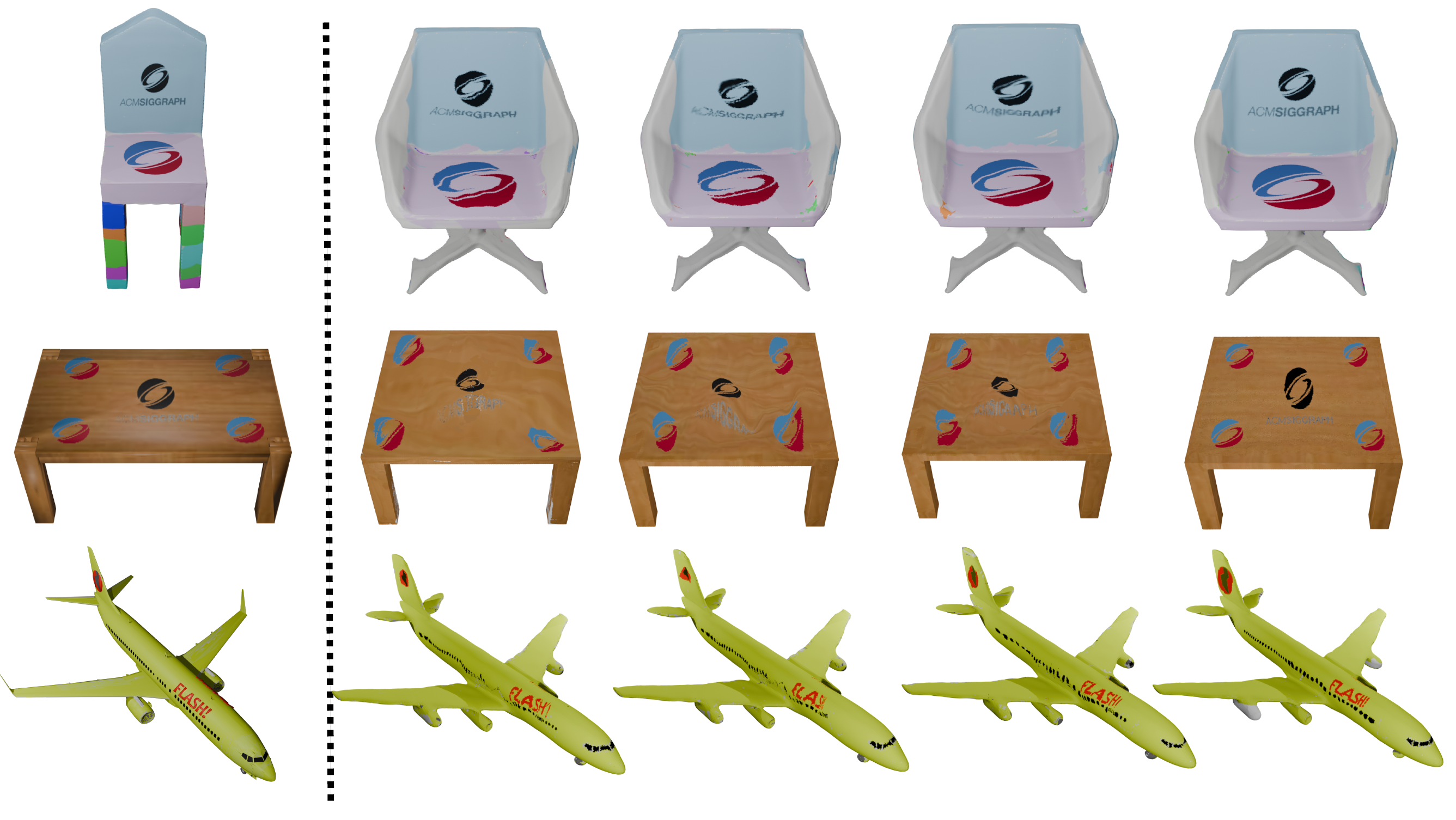}

    \put(4,56){\textbf{Source}}
    \put(30,56){\textbf{DIF}}
    \put(47,56){\textbf{DIT}}
    \put(65,56){\textbf{S-DIF}}
    \put(85,56){\textbf{Ours}}

\end{overpic}
\caption{\textbf{Texture transfer result in ShapeNet.}}
\label{Figure:Shape:texture_transfer_comparison}
\end{figure}
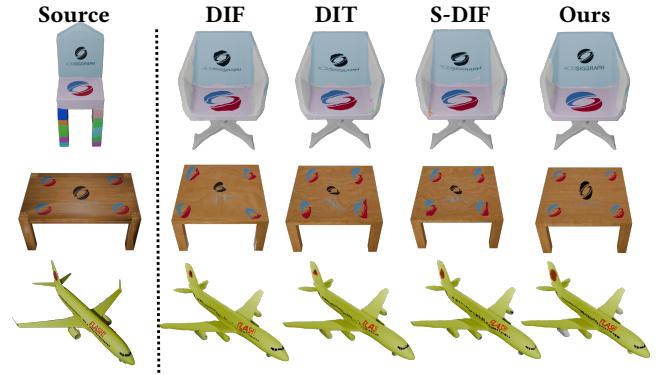

We compare texture transfer result using our method and baseline methods including DIF\cite{Deng2021DIF-Net}, DIT\cite{Zheng2021DIT-Net} and semantic DIF\cite{Kim2021SemanticDIF-Net}. A qualitative comparison is given in Figure \ref{Figure:Shape:texture_transfer_comparison}. Our results show better preservation in fine details and overall patterns of transfered texture compared to baseline approaches.

\section{Segmentation Granularity Ablation}
\label{Subsec:granularity_ablation}
\begin{figure}[h]
\captionsetup{skip=4pt}
\begin{overpic}[width=1.0\columnwidth]{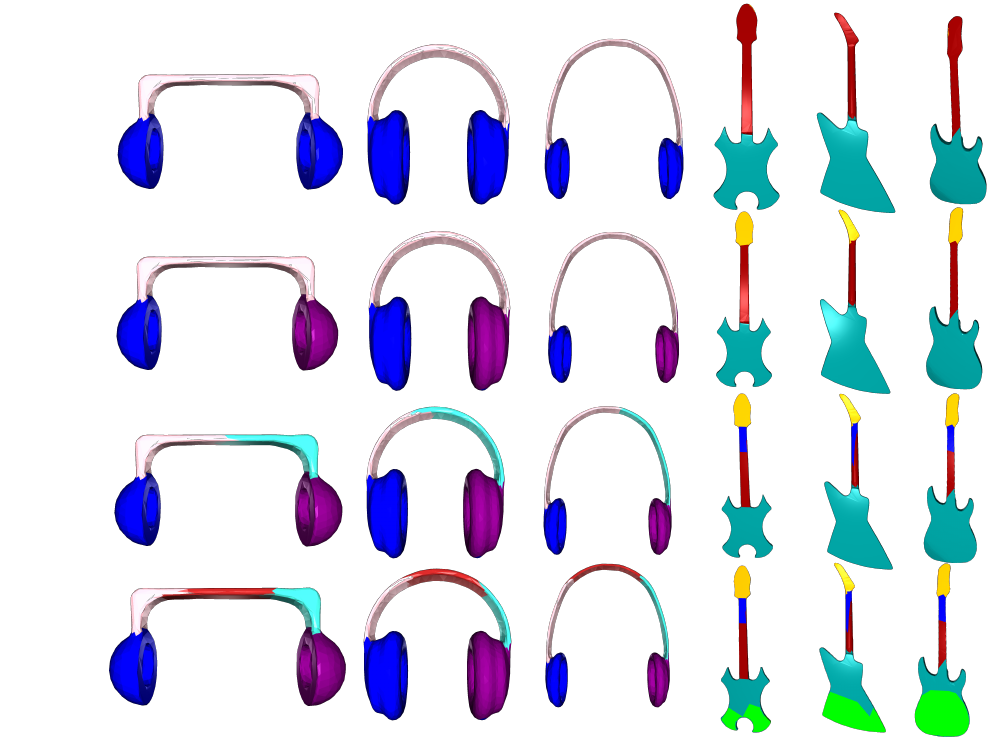}
    \put(1,60){\small{K=2}}
    \put(1,42){\small{K=3}}
    \put(1,24){\small{K-4}}
    \put(1,9){\small{K=5}}
\end{overpic}
\caption{\textbf{Ablation study on number of cluster K used in spectral consistent segmentation algorithm on earphone and guitar category}.}
\label{Figure:Shape:segmentation_granularity_ablation}
\end{figure}

We present segmentation result at different level of granularity in Figure ~\ref{Figure:Shape:segmentation_granularity_ablation} by using different number of cluster K in spectral consistent segmentation algorithm, as detailed in appendix \ref{App:SCS}. We achieve finer level segmentation by increasing number of clusters. In all other experiment, we adjust hyper-parameters of clustering so that the resulting segments are at the same level of granularity of baseline approaches.

\section{More qualitative results}
\begin{figure}[h]
\centering
\begin{overpic}[width=1.0\columnwidth]{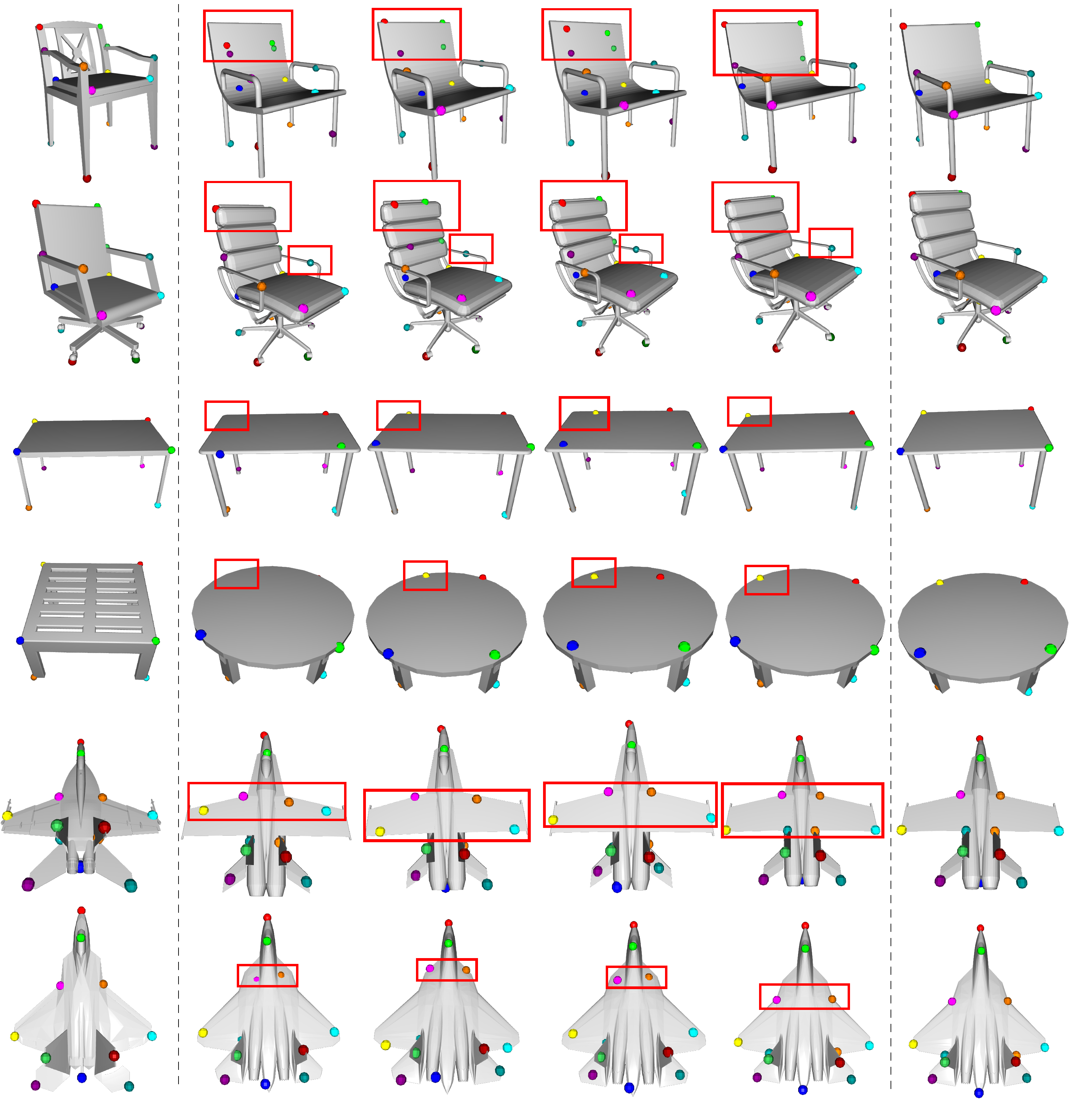}
    \put(5,101){\textbf{Source}}
    \put(21,101){\textbf{DIF}}
    \put(37,101){\textbf{DIT}}
    \put(51,101){\textbf{S-DIF}}
    \put(68,101){\textbf{Ours}}
    \put(85,101){\textbf{GT}}
\end{overpic}
\caption{\textbf{Additional comparison on keypoint transfer in ShapeNet.}}
\label{Figure:Shape:keypoint_transfer_supp}
\end{figure}

\begin{figure}[h]
\centering
\begin{overpic}[width=1.0\columnwidth]{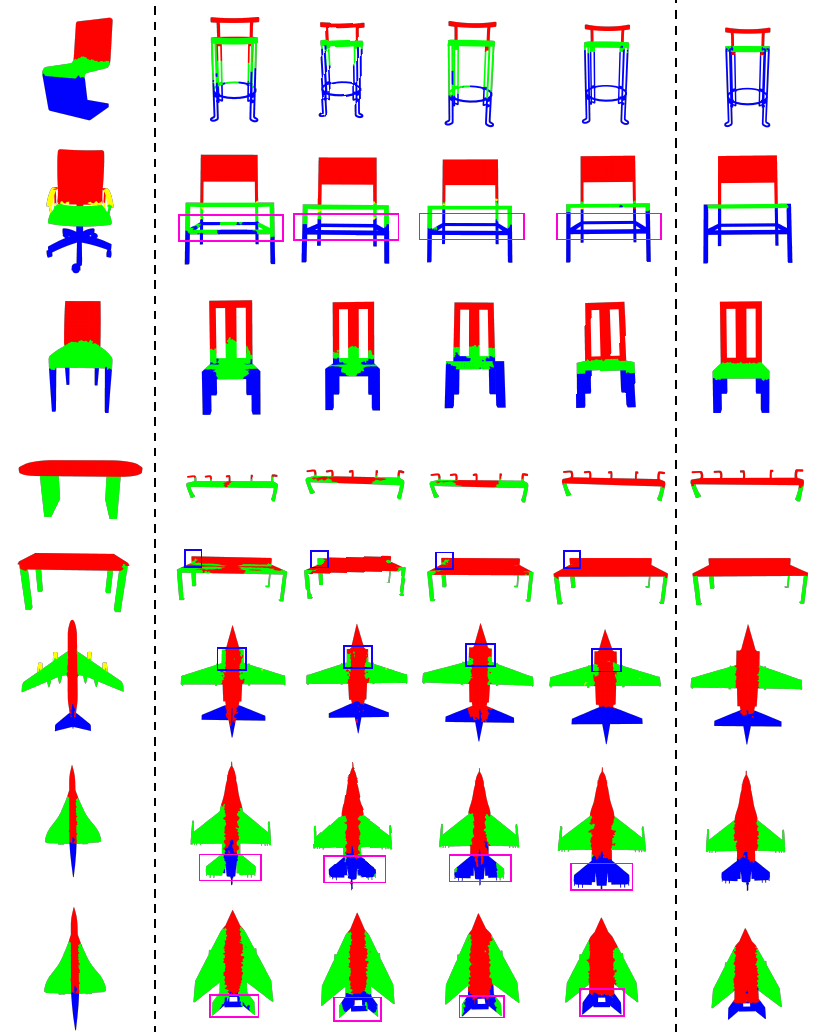}
    \put(6,100){\textbf{Source}}
    \put(21,100){\textbf{DIF}}
    \put(32,100){\textbf{DIT}}
    \put(44,100){\textbf{S-DIF}}
    \put(57,100){\textbf{Ours}}
    \put(71,100){\textbf{GT}}
\end{overpic}
\caption{\textbf{Additional comparison on label transfer in ShapeNet.}}
\label{Figure:Shape:label_transfer_supp}
\end{figure}

\begin{figure*}
\begin{overpic}[width=1.0\linewidth]{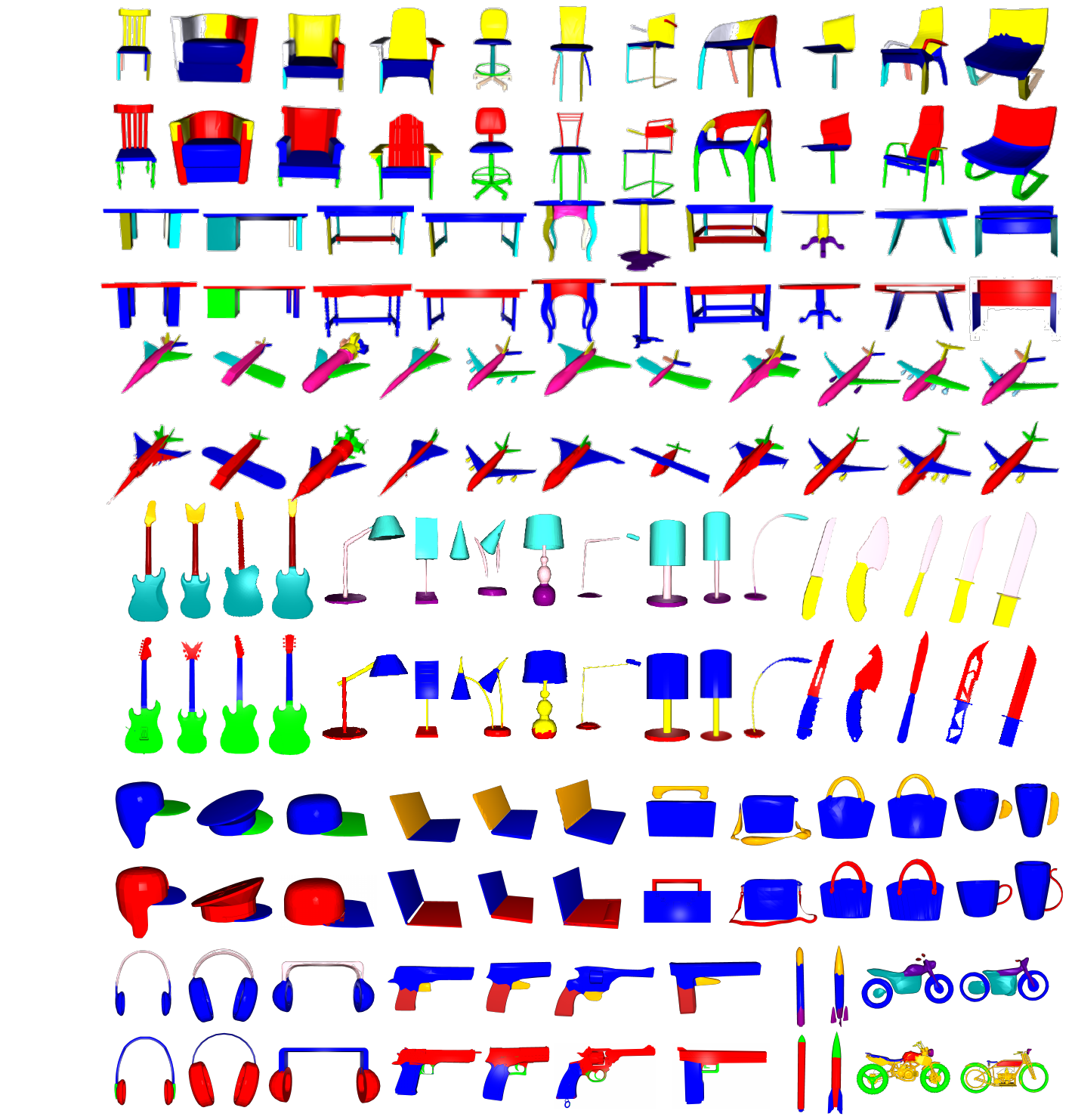}
    \put(1,95){\textbf{Ours}}
    \put(1,87){\textbf{GT}}
    \put(1,80){\textbf{Ours}}
    \put(1,73){\textbf{GT}}
    \put(1,66.5){\textbf{Ours}}
    \put(1,58){\textbf{GT}}
    \put(1,50){\textbf{Ours}}
    \put(1,38){\textbf{GT}}
    \put(1,27){\textbf{Ours}}
    \put(1,19){\textbf{GT}}
    \put(1,11.5){\textbf{Ours}}
    \put(1,3){\textbf{GT}}
    
\end{overpic}
\caption{\textbf{More qualitative results of shape co-segmentation in ShapeNet.}}    
\label{Figure:coseg_supp:More}
\end{figure*}
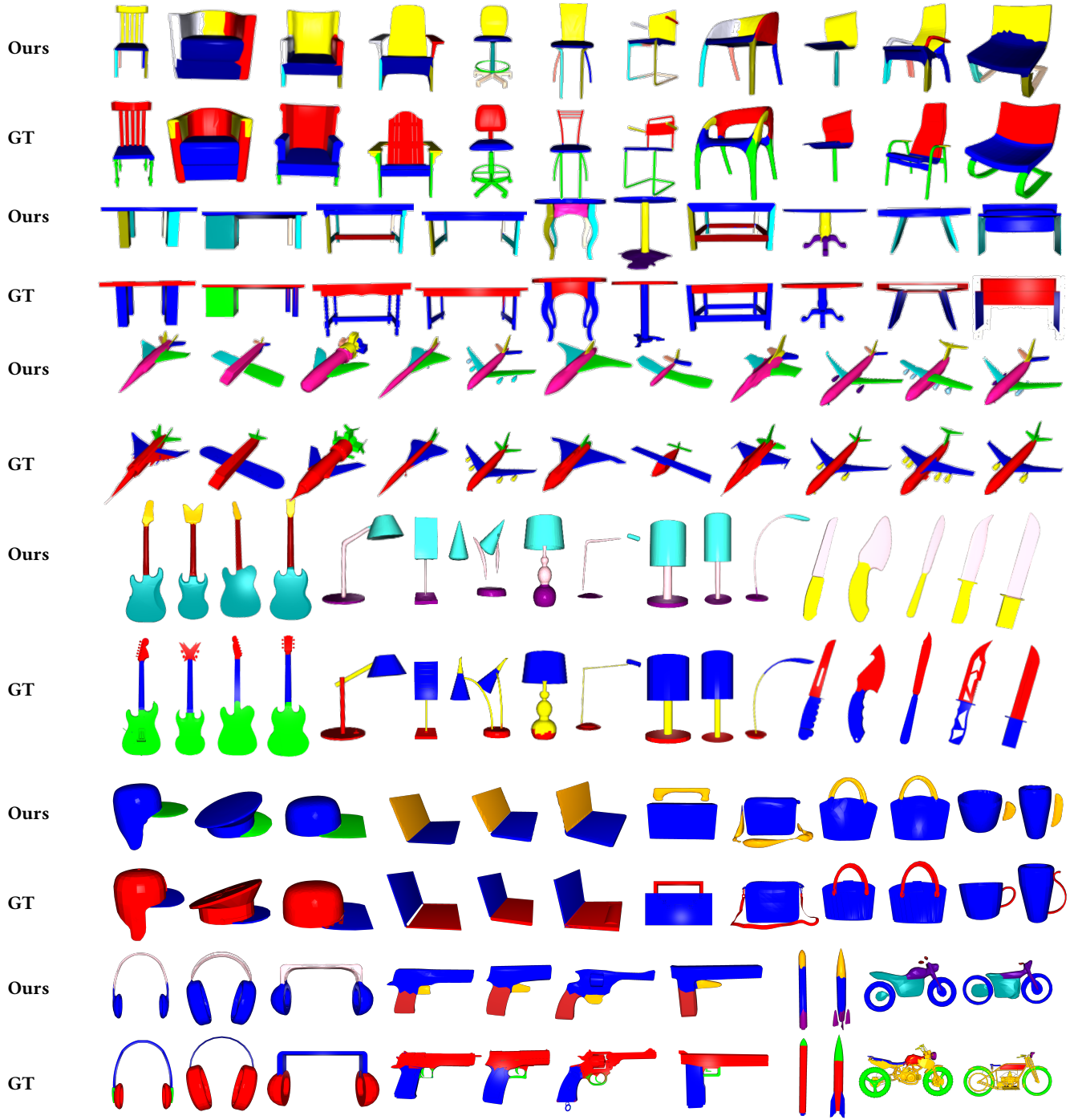

We show additional result on consistent segmentation, part label transfer and key point transfer in Figure  ~\ref{Figure:coseg_supp:More}, ~\ref{Figure:Shape:label_transfer_supp} and ~\ref{Figure:Shape:keypoint_transfer_supp}.

\newpage

\end{document}